\documentclass{aa}
\usepackage[varg]{txfonts}
\usepackage{natbib}
\usepackage{graphicx}
\usepackage{amsmath}
\usepackage{xcolor}
\usepackage[normalem]{ulem}
\bibpunct{(}{)}{;}{a}{}{,}

\begin{document}

\title{Asteroseismic sensitivity to internal rotation along the red-giant branch}
\author{F.\ Ahlborn\inst{1,2}
\and E.\ P.\ Bellinger\inst{3}
\and S.\ Hekker\inst{2,3}
\and S.\ Basu\inst{4}
\and G. C. Angelou\inst{1}}
\institute{Max-Planck-Institut f\"ur Astrophysik, Karl-Schwarzschild-Stra\ss e 1 ,85748 Garching, Germany\\
email: \url{fahlborn@mpa-garching.mpg.de}
\and Max-Planck-Institut f\"ur Sonnensystemforschung, Justus-von-Liebig-Weg 3, 37077 G\"ottingen, Germany
\and Stellar Astrophysics Centre, Department of Physics and Astronomy, Aarhus University, Ny
Munkegade 120, DK-8000 Aarhus C, Denmark
\and Department of Astronomy, Yale University, New Haven, CT 06520, USA
}
\date{Received xxx /
Accepted xxx }
\abstract
{Transport of angular momentum in stellar interiors is currently not well understood. Asteroseismology can provide us with estimates of internal rotation of stars and thereby advances our understanding of angular momentum transport.}
{We can measure core-rotation rates in red-giant stars and we can place upper bounds on surface-rotation rates using measurements of dipole ($l=1$) modes. Here, we aim to determine the theoretical sensitivity of modes of different spherical degree towards the surface rotation. Additionally, we aim to identify modes that can potentially add sensitivity at intermediate radii.}
{We used asteroseismic rotational inversions to probe the internal stellar rotation profiles in red-giant models from the base of the red-giant branch up to the luminosity bump. We used the inversion method of multiplicative optimally localised averages (MOLA) to assess how well internal and surface rotation rates can be recovered from different mode sets and different synthetic rotation profiles.}
{We confirm that dipole mixed modes are sufficient to set constraints on the average core-rotation rates in red giants. However, surface-rotation rates estimated with only dipole mixed modes are contaminated by the core rotation. We show that the sensitivity to surface rotation decreases from the base of the red-giant branch until it reaches a minimum at 60--80\% of the bump luminosity due to a glitch in the buoyancy frequency. Thereafter, a narrow range of increased surface sensitivity just below the bump luminosity exists. Quadrupole and octopole modes have more sensitivity in the outer parts of the star. To obtain accurate estimates of rotation rates at intermediate radii (i.e. a fractional radius of $\sim$0.4), acoustic oscillation modes with a spherical degree of $l\approx10$ are needed.}
{We find a minimum and subsequent maximum in the sensitivity to the surface rotation rate in red giants below the luminosity bump. Furthermore, we show that, if observed, quadrupole and octopole modes enable us to distinguish between differential and solid body rotation in the convection zone. This will be important when investigating the transport of angular momentum between the core and the envelope. }

\titlerunning{Inversions along the RGB}
\authorrunning{F.\ Ahlborn et al.}
\maketitle
\section{Introduction}

The rotation of a star has a substantial impact on its structure and evolution and has been shown to affect a number of internal processes. 
Rotation induces mixing in the stellar interior through mechanisms such as meridional flows and turbulence. 
This mixing brings fresh hydrogen into the stellar core, which then extends the main sequence lifetime of the star \citep[e.g.][]{eggenberger2010}.
However, the internal stellar rotation profile is not static, and adjusts considerably due to changes in the stellar structure, obeying local conservation of angular momentum. Additionally, the mixing processes also transport angular momentum throughout the stellar interior.

The hydrodynamical means of angular momentum transport ---meridional flows and turbulence---are not sufficient to explain the observed red-giant core rotation rates \citep{eggenberger2012,ceillier2013,marques2013,spada2016,eggenberger2017,ouazzani2019,fuller2019}. This implies that additional processes of angular momentum transport must be at work in stellar interiors. In order to remedy this situation, several studies have considered additional means. \cite{cantiello2014} considered the effects of magnetic fields on the internal rotation profile through the Tayler-Spruit (TS) dynamo \citep{spruit2002}. However, these latter authors found similarly insufficient angular momentum transport. \cite{fuller2014} explored the ability of internal gravity waves excited at the base of the convective envelope to transport angular momentum, and found that these waves indeed extract angular momentum from a fast rotating core; however, these waves only penetrate into the upper layers of the radiative zone on the red-giant branch (RGB), preventing them from slowing down the deep core on the RGB as observed by \cite{mosser2012} and \cite{gehan2018}. Subsequently, \cite{belkacem2015a,belkacem2015b} showed that mixed modes also efficiently remove angular momentum from the core in evolved red-giant models that are close to the luminosity bump. This is one potential explanation for the observed spin down of core rotation rates on the RGB. Most recently, \cite{fuller2019} modified the TS-dynamo and found larger magnetic field strengths which reproduce observed red giant core rotation rates.

To set more constraints on the internal processes that can transport angular momentum, detailed knowledge about the internal rotation profile of stars at different evolutionary stages is crucial. The deficiency in angular momentum transport becomes pronounced in the advanced evolutionary stages such as the subgiant and red-giant phases. 
Furthermore, the stellar oscillations that propagate through the interior of these stars provide an opportunity to probe their internal structure. Red giants exhibit solar-like oscillations that are stochastically excited by turbulent convection in the outer envelope. Due to the internal structure of red giants, many of these oscillation modes behave as gravity waves in the core and as pressure waves in the envelope. These modes are called mixed modes and probe the structure of red-giant cores \citep{beck2011,bedding2011,mosser2011,mosser2014}. Internal rotation splits non-radial oscillation modes into multiplets, where the frequency difference between multiplet components is called `rotational splitting'. The presence of rotationally split mixed modes facilitates the estimate of the core rotation rates in evolved stars. Measured rotational splittings in red giants indicate that the core rotates five to ten times faster than the surface \citep{beck2012,deheuvels2012}. Internal rotation rates as a function of depth have been estimated for a number of red giants using a forward modelling approach \citep{beck2014,beck2018} or by means of rotational inversions \citep{deheuvels2012,deheuvels2014,dimauro2016,triana2017,dimauro2018}. However, the exact shape of the internal rotation profile remains unknown, as only core and surface rotation rates can currently be measured. Rotational inversions probe core rotation relatively accurately, but the accuracy depends on the underlying rotation profile. Asteroseismic estimates of the surface rotation rate can only serve as upper limits given that the core rotates faster than the envelope. Rotation rates at intermediate radii are mostly unconstrained due to the lack of sensitivity of the observed oscillation modes to the internal rotation in these locations.

In this study, we perform rotational inversions in stellar models along the lower part of the RGB. We compute core- and surface-averaging kernels and investigate how their sensitivity to the core and surface rotation rate changes as the stellar models evolve. We interpret the changes in the sensitivities using the propagation diagrams of the stellar models. Subsequently, we select a single low-luminosity red-giant model to explore the prospect of using rotationally split quadrupole and octopole oscillation modes to probe the surface rotation in a red-giant model by means of asteroseismic rotational inversions. Up until now, only a few rotationally split $l=2$ modes have been measured in red-giant stars \citep{deheuvels2012,beck2014,dimauro2016}. Rotationally split $l\geq3$ modes have not been measured so far. Although modes of $l=2,3$ are expected to provide more sensitivity at the stellar surface, the measured $l=2$ modes have not yet been included in the red-giant rotational inversions. Here we use synthetic rotational splittings to demonstrate the diagnostic power of the $l=2,3$ modes for inferring the surface rotation rates of red giants. Finally, we extend the analysis to even higher spherical degrees ($l>3$) by analysing the lower turning points of the oscillation modes in order to demonstrate which modes probe rotation rates at intermediate radii.\\

\section{Synthetic data}
\begin{figure}
\resizebox{\hsize}{!}{\includegraphics{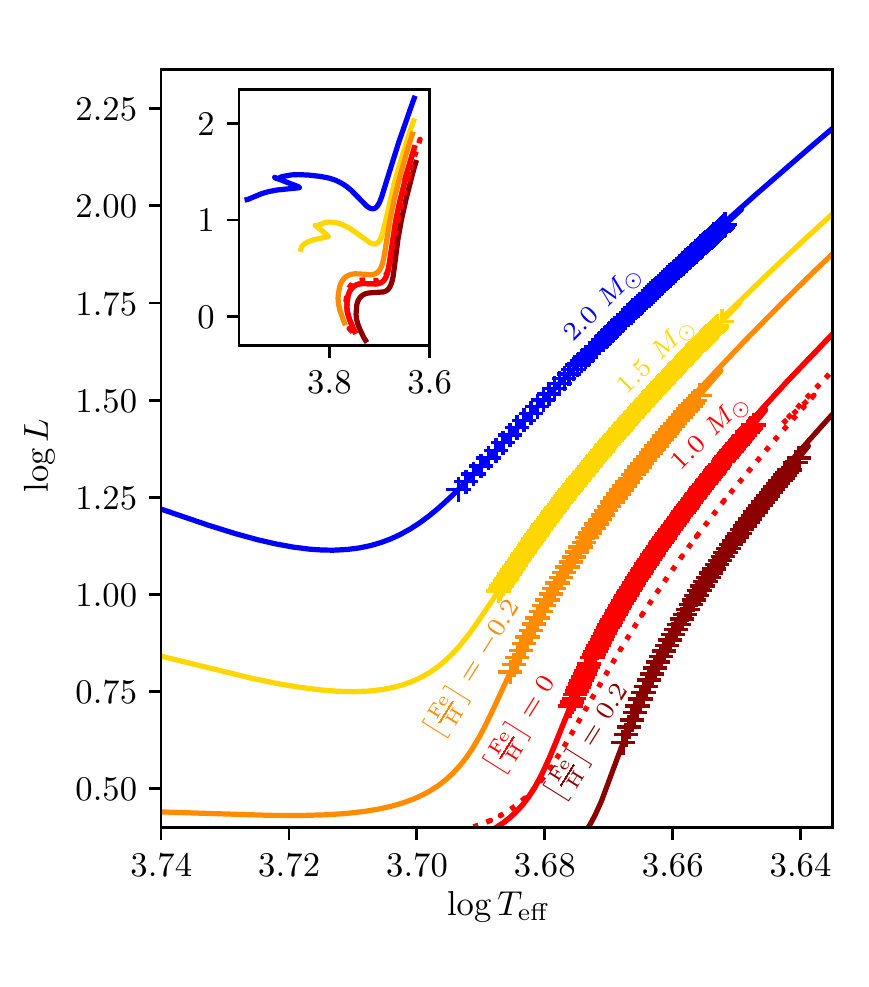}}
\caption{MESA evolutionary tracks for 1 (red), 1.5 (yellow), and 2~$M_\odot$(blue) models with solar metallicity are shown with solid lines in the respective colour. The evolutionary tracks for the 1~$M_\odot$ models with metallicities of $-0.2$ and 0.2 are shown with solid lines in orange and dark red respectively. The red dotted line indicates the GARSTEC 1~$M_\odot$ evolutionary track with solar metallicity. Models used for rotational inversions are marked with crosses in the respective colour. The inset shows the complete evolutionary tracks from the zero-age main sequence until beyond the bump in luminosity.}
\label{figevolmass}
\end{figure}
\label{secdata}
To study the characteristics of asteroseismic rotational inversions along the RGB we prepared different synthetic data sets comprising rotational kernels and rotational splittings.
We constructed a total of five different evolutionary tracks using Modules for Experiments in Stellar Astrophysics \citep[MESA, version 8845,][and references therein]{paxton2019} from the pre-main sequence to beyond the bump in the luminosity function.  We calculated three solar metallicity tracks with masses of $1.0,~1.5,$ and $2.0~M_\odot$ to explore the dependence of the inversion results on the stellar mass. The corresponding evolutionary tracks are shown in Fig.~\ref{figevolmass} in red, yellow, and blue. Additionally, we computed two evolutionary tracks with 1.0$~M_\odot$ and metallicities of $\left[\text{Fe/H}\right]=-0.2$ and $0.2$ to explore the dependence of the inversion results on the stellar metallicity. The corresponding evolutionary tracks are shown in Fig.~\ref{figevolmass} in orange and dark red. In addition, we computed another $1.0~M_\odot$ solar metallicity track using the GARching STellar Evolution Code \citep[GARSTEC,][]{weiss2008}. We selected the stellar parameter ranges such that solar-like oscillations are observable on the RGB. We computed the oscillation frequencies and corresponding rotational kernels of the models using the GYRE oscillation package \citep[][]{townsend2013,townsend2018}. 

We used the stellar models to compute synthetic data sets for rotational inversions. Rotational splittings were calculated for mode frequencies that lie in the range $\nu_\text{max}\pm2\Delta\nu$. For the study of inversion results along the RGB, we used mode sets containing modes with a spherical degree of $l=1$ only. We selected the most p-dominated and two of the least g-dominated modes per radial order in the given frequency range. For a detailed study of a single red-giant model, we also considered mode sets with spherical degrees $l\leq3$ and $l\leq10$. For spherical degrees $l>1$, only the most p-dominated modes were selected. To compute synthetic rotational splittings we imposed different synthetic rotation profiles. We chose two step profiles with the step at the base of the convective envelope and at 1.5 times the radius of the hydrogen-burning shell. We refer to these profiles as the `envelope step' and `core step profile' respectively in the rest of this paper. Further, we defined a profile with constant rotation below the base of the convection zone and a power-law decrease in the convective envelope. We refer to this profile as the `convective power law profile'. In addition to the rotational splittings, we also computed synthetic uncertainties. We used a quadratic model for the uncertainties following \cite{schunker2016a}. The uncertainty model was calibrated to observational data from a low-luminosity red giant and then shifted to the frequency range of the respective stellar model.
A detailed description of the stellar models and the synthetic data sets can be found in Appendix~\ref{secdataapp}.

\section{Rotational inversions}
\label{secrotinv}
Stellar rotation splits oscillation modes with spherical degrees of $l>0$ into multiplets of $2l+1$ modes. These modes are denoted by the azimuthal order $m$ with ${-l\leq m\leq l}$. The frequency difference between modes of subsequent azimuthal orders, that is, the rotational~splitting, is denoted by~$\delta\omega_{nl}$. Here, $n$ refers to the radial order of the oscillation mode. For slow rotation, the rotational splitting is a first-order perturbation to the mode frequencies $\omega_{nl}$:
\begin{align}
    \omega_{nlm}=\omega_{nl}+m\,\delta\omega_{nl,}
\end{align}
where $\delta\omega_{nl}$ is given as:
\begin{align}
    \delta\omega_{nl}=\int_0^R\mathcal{K}_{nl}(r)\Omega(r)\,\text{d}r,
    \label{eqsplitting}
\end{align}
using the angular velocity profile $\Omega(r)$ and so-called `rotational kernels' $\mathcal{K}_{nl}(r)$ \citep[e.g.][]{gough1981}, which can be calculated from a stellar model. To compute the internal stellar rotation profile from the rotational splittings, we used the multiplicative optimally localised averages (MOLA) inversion technique \citep{backus1968}. The idea of MOLA inversions is to linearly combine rotational kernels to form so-called `averaging kernels' $K(r,r_0)$ which are, in an ideal case, localised at their target radius $r_0$:
\begin{align}
K(r,r_0)=\sum_{i\in \mathcal{M}}c_i(r_0)\,\mathcal{K}_i(r),
\end{align}
where $\mathcal{M}$ is the mode set of interest and $i$ refers to a combination of $(n,l)$. The inversion coefficients $c_i(r_0)$ have to obey the constraint of unimodular averaging kernels \citep[e.g.][]{backus1968}. Provided that the averaging kernels are well localised at the given target radius the rotation rate can be probed at $r_0$:
\begin{align}
    \bar{\Omega}(r_0)=\int_0^RK(r,r_0)\Omega(r)\,\text{d}r=\sum_{i\in\mathcal{M}}c_i(r_0)\,\delta\omega_i,
    \label{eqomega}
\end{align}
where the definition of the rotational splittings from Eq.~(\ref{eqsplitting}) has been used. The estimated rotation rate $\bar{\Omega}(r_0)$ is an average value with the averaging kernel acting as the weighting function. 

To analyse the results of the rotational inversions, it is convenient to split the estimated rotation rates $\bar\Omega$ into a core ($r\leq r_\text{core}$) and an envelope ($r>r_\text{core}$) contribution, where $r_\text{core}$ does not need to coincide with the actual stellar core radius. Assuming a step-like rotation profile with a step at $r_\text{core}$, the two parts of the integral in Eq.~(\ref{eqomega}) can be solved, leading to
\begin{align}
    \bar{\Omega}(r_0)=\beta_\text{core}(r_0)\cdot\Omega_\text{c}+(1-\beta_\text{core}(r_0))\cdot\Omega_\text{e},
    \label{eqbeta}
\end{align}
with 
\begin{align*}
    \beta_\text{core}(r_0)=\int_0^{r_\text{core}}K(r,r_0)\,\text{d}r\,.
\end{align*}
Another useful quantity for the analysis of rotational inversion results is the cumulative averaging kernel, which is the integral of the averaging kernel as a function of the radius. This kernel shows in a straightforward manner the part of the star that the averaging kernel is most sensitive to.

To compute localised averaging kernels, the inversion coefficients $c_i$ are chosen such that both the radial spread in the averaging kernels and the errors due to data uncertainties are simultaneously small. This is implemented by minimising the objective function:
\begin{align}
    \int_0^RK(r,r_0)^2J(r,r_0)\,\text{d}r+\frac{\mu}{\mu_0}\sum_{i,j\in\mathcal{M}}c_i(r_0)c_j(r_0)E_{ij},
    \label{eqobj}
\end{align}
where $E_{ij}$ denotes the covariance matrix of the measured rotational splittings. Here, $J(r,r_0)$ is a function which vanishes at $r_0$ and otherwise monotonically increases. A common choice is $J(r,r_0)=12(r-r_0)^2/R$ \citep[][]{backus1968,gough1985}. The so-called trade-off parameter $\mu$ balances the uncertainty on the solution because of propagated errors and the resolution of the inversions as indicated by the width of the averaging kernels. The quantity $\mu_0$ is defined as
\begin{align}
    \mu_0=\frac{1}{N}\sum_{i\in\mathcal{M}}E_{ii},
\end{align}
where $N$ denotes the number of modes. This choice ensures that the second part of the objective function (Eq.~\ref{eqobj}) is dimensionless and the inversion results do not depend on the absolute values of the measurement uncertainties \citep[e.g.][]{gough1985}. The uncertainties on the estimated rotation rates $\bar{\Omega}$ are calculated through error propagation from Eq.~(\ref{eqomega}).

\section{Inversion results along the RGB}
\label{secinvRGB}
In this section, we study how rotational inversion results change as stars evolve along the RGB.
We performed rotational inversions for our stellar models between the base of the RGB and the luminosity bump (marked regions in Fig.~\ref{figevolmass}).

For each selected model we computed a core (${r_0/R=0.003}$) and a surface (${r_0/R=0.98}$) averaging kernel based on the synthetic data sets described in Sect.~\ref{secdata} and Appendix~\ref{secdataapp}. We used these averaging kernels to compute core and surface rotation rates given the synthetic rotational splittings. The trade-off parameter was tuned as described in Appendix~\ref{sectradeoff}. Here, we are specifically interested in the sensitivities of the resulting averaging kernels\footnote{We would like to point out that the localisation of the averaging kernels does not depend on the rotational splittings according to Eq.~(\ref{eqobj}), but only on data errors, the number of modes available, and the region probed by the modes used in the mode set.}.  In the following, we therefore focus on the sensitivity of the averaging kernels to different regions inside the star. To study the sensitivity of the computed averaging kernels to the region of interest, we split them into two zones according to Eq.~(\ref{eqbeta}). We calculated the sensitivity of the core-averaging kernel in the core $\beta_\text{core}(0.003R)$ and the sensitivity of the surface-averaging kernel in the envelope ${\beta_\text{surf}=1-\beta_\text{core}(0.98R)}$. We computed the sensitivities for different values of $r_\text{core}=r_\text{rcb},1.5r_\text{H},0.1R,0.2R$. Choosing the base of the convection zone ($r_\text{rcb}$) or the hydrogen burning shell ($r_\text{H}$) as the integration boundaries takes structural changes into account whereas choosing a fixed value in fractional radius ignores structural changes. The results appear qualitatively similar for all separations between core and envelope.
\begin{figure}
\resizebox{\hsize}{!}{\includegraphics{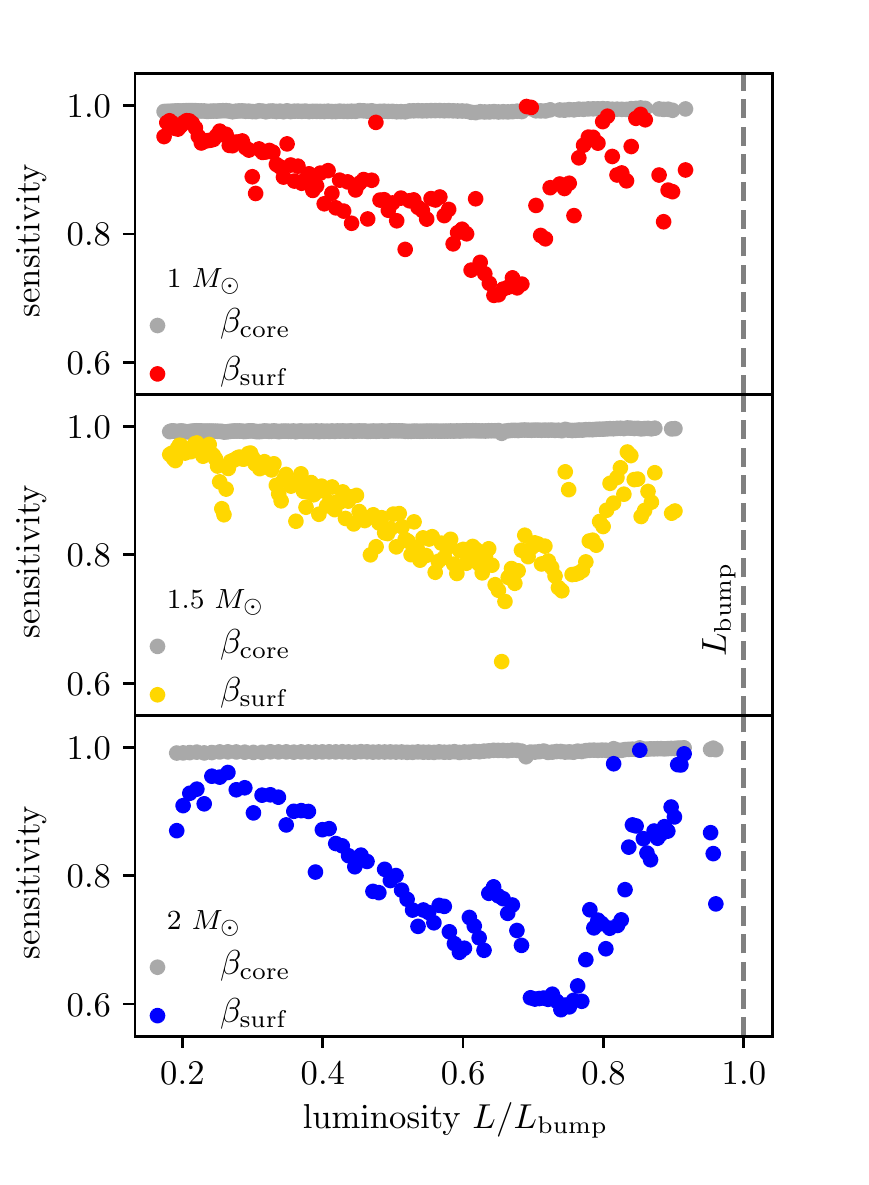}}
\caption{Core and surface sensitivities $\beta_\text{core}$ and $\beta_\text{surf}$ for the solar metallicity evolutionary tracks with 1, 1.5 and 2~$M_\odot$.}
\label{figsensMESA}
\end{figure}
\begin{figure}
\resizebox{\hsize}{!}{\includegraphics{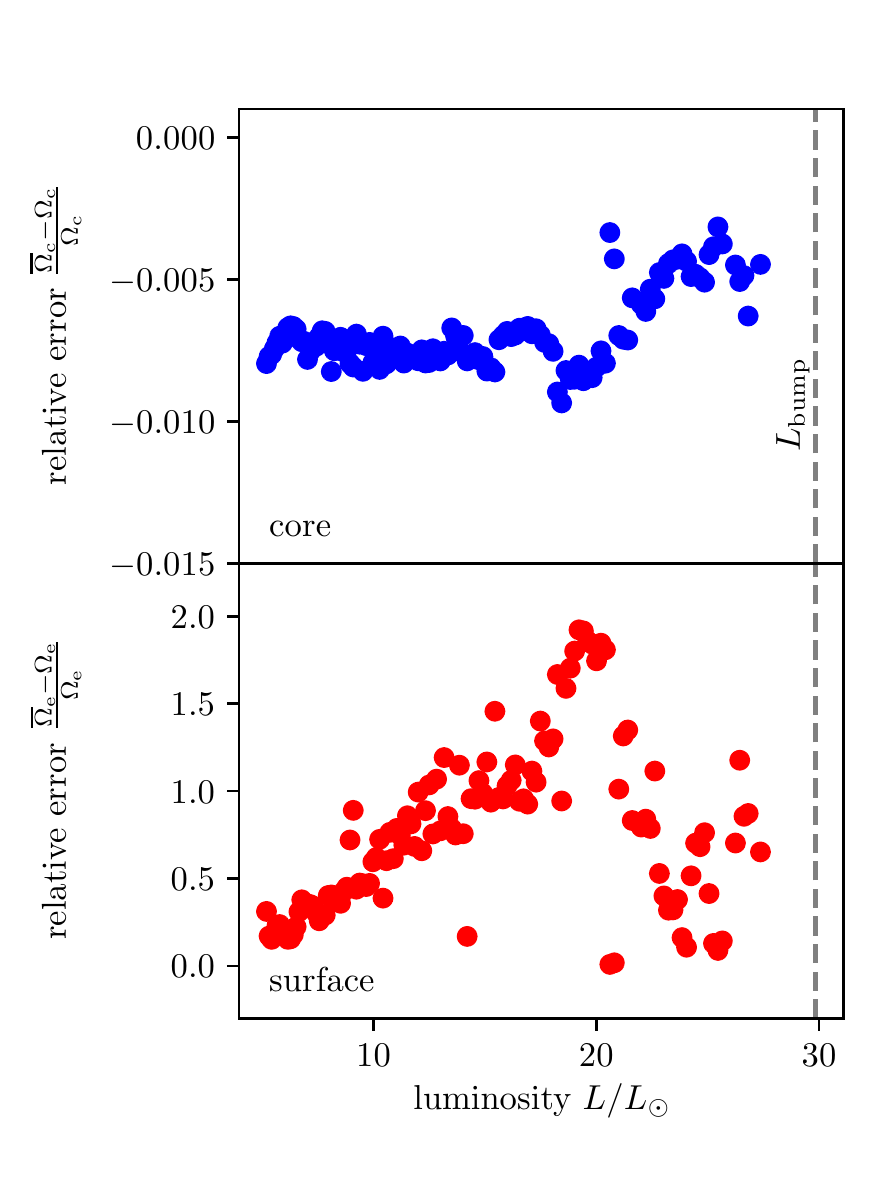}}
\caption{Relative errors of core and surface rotation rates along the RGB for the $1~M_\odot$, $[\text{Fe}/\text{H}]=0$ MESA models. Here the envelope step profile has been used. (See Appendix~\ref{secrotprofiles} for the description of the rotation profiles.)}
\label{figinvrgbMESA}
\end{figure}
\subsection{Core and surface sensitivities}
The rotational inversion results in terms of core and surface sensitivities $\beta_\text{core}$ and $\beta_\text{surf}$ for the three different stellar masses are shown in Fig.~\ref{figsensMESA} as a function of the stellar luminosity. The sensitivities for stellar models with varying metallicity are shown in Fig.~\ref{figsensmetal}. Here we choose $r_\text{core}=r_\text{rcb}$. Using the estimated rotation rates and the known input rotation profiles we compute relative errors for the rotation rates. Figure~\ref{figinvrgbMESA} shows the relative errors of the estimated rotation rates for the 1~$M_\odot$, $[\text{Fe}/\text{H}]=0$ track as a function of luminosity using the envelope step rotation profile. 

The grey symbols in Fig.~\ref{figsensMESA} show that the sensitivity of the core-averaging kernel is confined below the base of the convection zone in its entirety. This means that the core-averaging kernels are well localised. Accordingly, relative errors in core rotation rates are small all along the RGB. As shown in the upper panel of Fig.~\ref{figinvrgbMESA} core rotation rates can be recovered with a relative error of less than 1\%. The relative error is negative because some of the sensitivity of the core-averaging kernel is located in the slowly rotating envelope. According to Eq.~(\ref{eqomega}), this decreases the estimated core rotation rate. We conclude that a set of 12 dipole modes is sufficient to recover core rotation rates given the envelope step profile.
The inversion results for core rotation rates appear similar as in Fig.~\ref{figinvrgbMESA} for other synthetic rotation profiles as long as the width of the fast-rotating core region is equal to or larger than the width of the core-averaging kernel.

The inversion results for the surface-averaging kernels (Fig.~\ref{figsensMESA}, coloured symbols) show a very different behaviour. As the star evolves up the RGB, the surface-averaging kernel becomes increasingly sensitive to the core region. At about $L=0.6-0.8~L_\text{bump}$, the surface sensitivity $\beta_\text{surf}$ reaches a minimum of 70\%, which translates to 30\% sensitivity to the core rotation. Here, $L_\text{bump}$ indicates the bump luminosity.  The relative error of the surface rotation rate as a function of luminosity (Fig.~\ref{figinvrgbMESA}, lower panel) reflects the behaviour of the sensitivities; it reaches a maximum of about 200\% which coincides with the minimum in sensitivity. This translates to an absolute error of 200~nHz given the imposed rotation profile. After that maximum, the relative error decreases again. In this case, relative errors are positive because the estimated surface rotation rate is contaminated by the fast core rotation. We note that even at low luminosities around 5-7~$L_\odot$ the relative error amounts to about 20\%.

Although a few models have surface averaging kernels which are almost insensitive to the core, in general a set of 12 dipole modes yields estimated surface rotation rates which suffer from systematic errors. We have found the same qualitative behaviour in the inversion results for all our evolutionary tracks shown in Fig.~\ref{figevolmass}. We therefore conclude that the variation of the surface sensitivity is independent of mass and metallicity in the ranges we investigated.

The scatter of the computed sensitivities in Fig.~\ref{figsensMESA} most probably results from slightly different sensitivities of the individual modes selected in subsequent models. Given the relatively small scatter, we think that the minimum in the sensitivity curves is a physical feature of the models. To rule out that the calibration of the trade-off parameter has an impact on the inversion results we recomputed inversion results for a fixed value of the trade-off parameter $\mu=0$. The inversion results appeared very similar to results where we tuned the trade-off parameter.

\begin{figure}
\resizebox{\hsize}{!}{\includegraphics{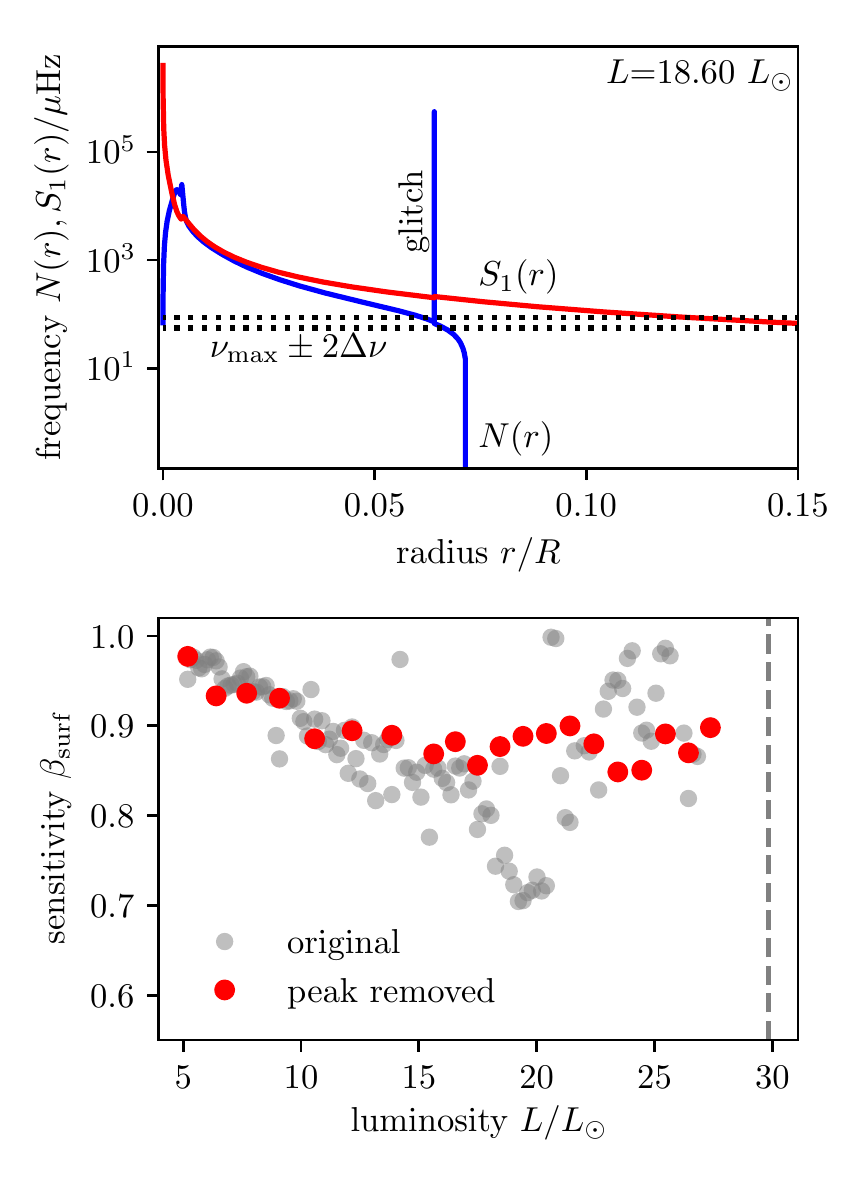}}
\caption{Upper panel: Propagation diagram for a 1.0~$M_\odot$ model in the vicinity of the surface-sensitivity minimum. Here, $N$ indicates the buoyancy frequency and $S_1$ indicates the Lamb frequency for $l=1$ modes. The frequency range of the dipole modes used for the rotational inversions is indicated with two horizontal dotted lines. Lower panel: Surface sensitivity as a function of stellar luminosity. Red dots indicate inversion results for models where the peak associated with the composition discontinuity has been removed manually from the buoyancy frequency. Grey dots show the original inversion results from the upper panel of Fig.~\ref{figsensMESA}  for comparison.}
\label{figprop}
\end{figure}

To understand what potentially causes this decrease in sensitivity we studied the propagation diagrams of the stellar models and how they change with stellar evolution.  A propagation diagram is shown in the upper panel of Fig.~\ref{figprop} for a 1.0~$M_\odot$ model ascending the RGB with a luminosity of $L$=18.60~$L_\odot$, which is right at the minimum of the sensitivities. A sharp peak is visible in the buoyancy frequency which is caused by the composition discontinuity left behind by the convective envelope. In Fig.~\ref{figprop} the observable frequency range just crosses the  `base' of the peak in the buoyancy frequency. Before that point, the peak in the buoyancy frequency lies within the frequency range of the modes used for the rotational inversions. We suggest that the peak enhances the coupling between the p- and the g-modes. \cite{cunha2015} showed that the combined effect of mode coupling and a buoyancy glitch leads to an increase of the period spacing for the p-dominated mixed modes. This means that the oscillation modes exhibit a more g-like behaviour and are then more sensitive to the core. After the minimum in sensitivity, the peak moved out of the frequency range and the coupling decreases again. Oscillation modes become more sensitive to the surface again. The point in evolution where the observable frequency range crosses the base of the buoyancy glitch also coincides with the minimum in sensitivities for the other models under consideration.

To study the effect of the glitch in the buoyancy frequency in more detail we artificially removed the glitch in a couple of stellar models from the 1~$M_\odot$, $\left[\text{Fe}/\text{H}\right]=0$ track and recomputed the rotational inversions in the same way as before. The surface sensitivities for these modified models are shown in the lower panel of Fig.~\ref{figprop} in red. The original inversion results from the upper panel of Fig.~\ref{figsensMESA}  are shown in in grey for comparison. Models without the peak in the buoyancy profile behave substantially differently from the original results. The minimum in surface sensitivities around 19~$L_\odot$ is no longer present. Instead, the surface sensitivity decreases steadily when the models evolve along the RGB. This indicates indeed that it is the glitch in the buoyancy profile that leads to the occurrence of a  minimum and subsequent maximum in surface sensitivities.
\subsection{Further tests}
We also computed relative errors assuming two other synthetic rotation profiles and found similar results as for the envelope step profile; see Figs~\ref{figrelerrcorestep} and \ref{figrelerrrpowerlaw} in the Appendix for details. Differences among the relative errors for different synthetic rotation profiles can be completely explained by the combination of the sensitivities of the averaging kernels and the synthetic rotation profiles according to Eq.~(\ref{eqomega}).

We performed the same analysis on the GARSTEC $1.0~M_\odot$ model. The mode selection and computation of synthetic data were done in the same way as before. The inversion results are shown in Fig.~\ref{figinvrgbGARS}. The results appear very similar to those computed for the MESA stellar evolution code. The amplitude of the increase in relative errors and the location of the maximum are particularly similar. We therefore find our conclusions to be independent of the  code used.

\section{Inversion results with $l\geq1$}
\label{secres}
Previous studies have shown that red giant rotation rates cannot be well constrained at intermediate radii using rotational inversions \citep{deheuvels2012,deheuvels2014,dimauro2016,triana2017}. We therefore first focus on core and surface rotation rates and subsequently address the intermediate radii.

From the 1.0~$M_\odot$ $[\text{Fe}/\text{H}]=0$ MESA track we selected one model with $L=7.67~L_\odot$ and $T_\text{eff}=4690~\text{K}$ for a more detailed study. This is an early low-luminosity red-giant model. The stellar parameters of this model are similar to previously studied stars \citep[e.g.][]{deheuvels2012,deheuvels2014,dimauro2016,triana2017}. 
\subsection{Core rotation}
\begin{figure}
\resizebox{\hsize}{!}{\includegraphics{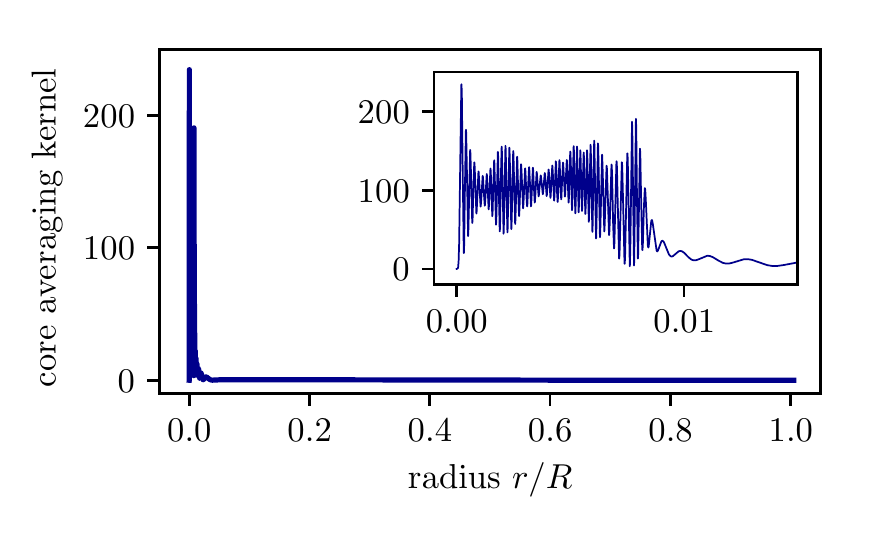}}
\caption{Core-averaging kernel obtained with the $l=1$ mode set. The inset shows an enlarged view of the averaging kernel in the core.}
\label{figcoreaverage}
\end{figure}

\begin{figure}
\resizebox{\hsize}{!}{\includegraphics{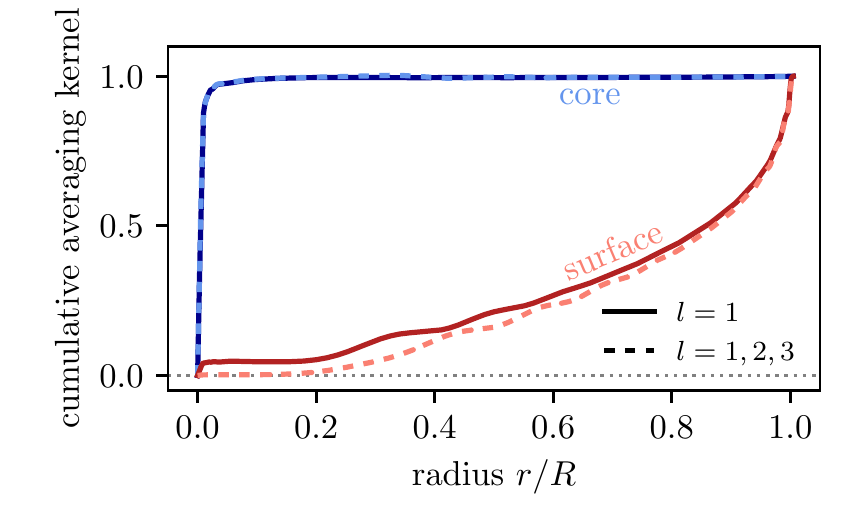}}
\caption{Cumulative core- and surface-averaging kernels for the $l=1$ and the $l=1,2,3$ mode set are shown with solid and dashed lines respectively. The blue curves indicate the core-averaging kernel, while the red curves indicate the surface-averaging kernel.}
\label{figl1cumul}
\end{figure}

We calculate core rotation rates using an averaging kernel localised at $r_0/R=0.003$ (Fig.~\ref{figcoreaverage}). The core-averaging kernel has a strong peak in the core and is otherwise flat. The inset in Fig.~\ref{figcoreaverage} shows some additional small-scale oscillations in the core region arising from the small-scale oscillations in the rotational kernels. These small-scale oscillations do not affect our ability to measure rotation rates as long as the true rotation profile changes only on scales larger than these oscillations. Furthermore, the cumulative averaging kernel in Fig.~\ref{figl1cumul} (blue solid line) rises steeply at very low radii, despite the small-scale oscillations. This shows that its sensitivity is confined to very low radii. 


The core rotation rates calculated with the $l=1$ mode set using the three different synthetic rotation profiles as described in Appendix~\ref{secrotprofiles} are listed in the upper part of Table~\ref{tablel1}. For the core step profile, the estimated core rotation rate is significantly different from the input rotation rate. This difference occurs because the core-averaging kernel shows about $10\%$ of its sensitivity outside the fast rotating core (which corresponds to $\beta_\text{core}=0.9$). For the other two rotation profiles, the estimated core rotation rates match the input within the $1\sigma$ uncertainties. In these cases, the core-averaging kernel shows less leakage outside the fast rotating region.

The estimated core rotation rates improved only marginally when the quadrupole and octopole modes (Table \ref{tablel1}, lower part) were included. In particular, the estimated core rotation rate for the core step rotation profile remained substantially different from the input as the core-averaging kernel  still shows about $10\%$ of its sensitivity outside $1.5r_\text{H}$. The cumulative core-averaging kernel (Fig.~\ref{figl1cumul}, blue dashed line) derived from the $l=1,2,3$ mode set closely resembles that obtained using only the dipole modes. 

\begin{table}
\caption{Estimated core and surface rotation rate for the different synthetic rotation profiles. The definitions of the synthetic rotation profiles are given in Appendix~\ref{secrotprofiles}.}
\label{tablel1}
\centering
\begin{tabular}{c c c c}
\hline\hline
\rule{0pt}{12pt}   &  &Core& Surface\\
Mode set&Profile&rotation&rotation\\
&&$[\text{nHz}]$&$[\text{nHz}]$\\[.5ex]
\hline
\rule{0pt}{10pt}$l=1$& core step & $686\pm6 $ & $128\pm7$ \\
& envelope step & $745\pm6$ &$129\pm7$\\
& convective power law&$747\pm6$ & $179\pm7$\\
\hline
\rule{0pt}{10pt}$l=1,2,3$& core step & $687\pm6$ & $101\pm4$ \\
& envelope step & $745\pm6$ &$102\pm4$\\
& convective power law&$748\pm6$ & $151\pm4$\\
\hline
\rule{0pt}{10pt}input& ... & $750$ &$100$ \\
\hline
\end{tabular}
\tablefoot{The inversion results were calculated using the $l=1$ mode set and the $l=1,2,3$ mode set. Core and surface rotation rates were calculated with target radii of $r_0/R=0.003$ and $r_0/R=0.98,$ respectively. The last row contains the input values for core and surface rotation rates in the different profiles.}
\end{table}

\begin{figure}
\resizebox{\hsize}{!}{\includegraphics{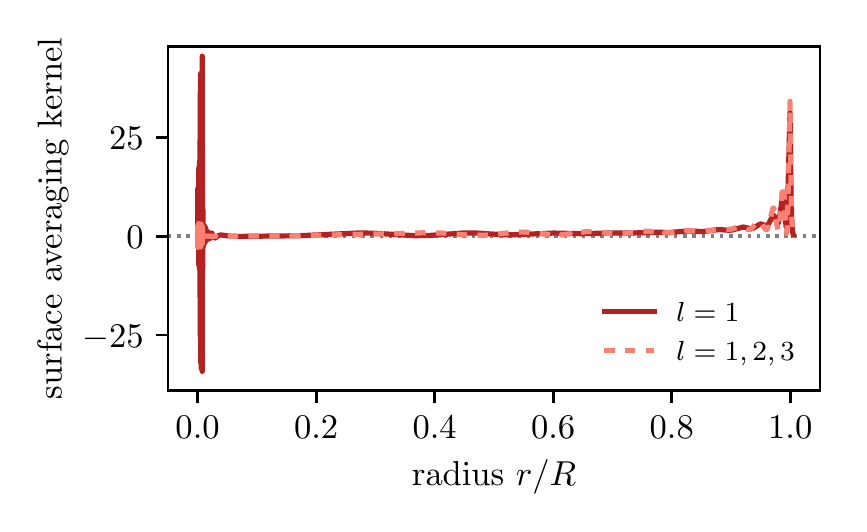}}
\caption{Surface-averaging kernels for different mode sets. The $l=1$ mode set is shown with the solid line and the $l=1,2,3$ mode set is shown with the dashed line.}
\label{figaverage}
\end{figure}

\subsection{Surface rotation}
We estimated surface rotation rates with an averaging kernel localised at $r_0/R=0.98$. While the surface-averaging kernel for the $l=1$ mode set (Fig.~\ref{figaverage}, red solid line) shows more sensitivity towards the surface than the core-averaging kernel, it is not well localised and shows sensitivity throughout the whole star. This behaviour is also visible in the cumulative surface-averaging kernel (Fig.~\ref{figl1cumul}, red solid line), confirming that this surface-averaging kernel is not localised at the target radius as intended. This means that the estimated surface rotation rate is not just sensitive to the stellar surface rotation, but also to the rotation rate in the deeper layers. According to Eq.~(\ref{eqomega}), the estimated rotation rate is rather an average value through the whole stellar envelope and the stellar core.

The estimated surface rotation rates for the $l=1$ mode set listed in the upper part of Table~\ref{tablel1} differ significantly from the input values for all three synthetic profiles. The surface-averaging kernel shows about $5\%$ of its sensitivity below $1.5r_\text{H}$ and $r_\text{rcb}$. We note that Eq.~(\ref{eqbeta}) tells us that the sensitivity in the deep core increases the estimated surface rotation rate provided that the core rotates significantly faster than the envelope. For the convective power law profile, the estimated surface rotation rate differs even more from the input value. Due to the large spread of the surface-averaging kernel, a large range in radius contributes to the integral in Eq.~(\ref{eqomega}). Since the rotation rate for any point below the surface is higher than at the surface in the convective power law rotation profile, the estimated surface rotation rate is substantially larger than the true value and serves only as an upper limit. \cite{dimauro2016} also estimated core and surface rotation rates in a low-luminosity red giant using alternative methods to inversions. These latter authors find that estimated core rotation rates agree among the different methods within the given $1\sigma$-uncertainties, but that estimated surface rotation rates on the other hand do not agree within the given uncertainties and vary over a large range of values. This indicates that a method that can provide a reliable estimate of the surface rotation using dipole modes only has not yet been developed.

The inclusion of quadrupole and octopole p-dominated mixed modes improves the estimated surface rotation rates considerably. The estimated surface rotation rates for the $l=1,2,3$ mode set (Table~\ref{tablel1}, lower part) now match the input rates within the $1\sigma$ uncertainties for both the core and envelope step rotation profile. The surface-averaging kernel shows only less than $1\%$ of its sensitivity below $1.5r_\text{H}$ and $r_\text{rcb}$ (Fig.~\ref{figaverage}, red dashed line). This is reflected by the cumulative surface-averaging kernel (Fig.~\ref{figl1cumul}, red dashed line) which does not show any net sensitivity in the core, that is, below $r_\text{rcb}=0.129R$. The calculated surface rotation rate is therefore almost independent of the stellar core rotation, such that the estimated value is an average rotation rate from the base of the convection zone outwards. The estimate of the surface rotation for the convective power law profile also improved considerably. However, it still does not match the input rotation rate to within $1\sigma$ uncertainty. This arises for the same reasons as that for the $l=1$ mode set. To get an accurate estimate of the surface rotation assuming this rotation profile the resolution of the surface-averaging kernel needs to be increased even further. This can be only achieved by including higher degree modes ($l>3$) as in solar rotational inversions \citep[e.g.][]{christensen1990}.  We calculated also rotational inversions for mode sets containing $l=1,2$ and $l=1,3$ modes. These tests showed that mainly the $l=3$ modes are responsible for the decrease in sensitivity to the core rotation.

\begin{figure}
\resizebox{\hsize}{!}{\includegraphics{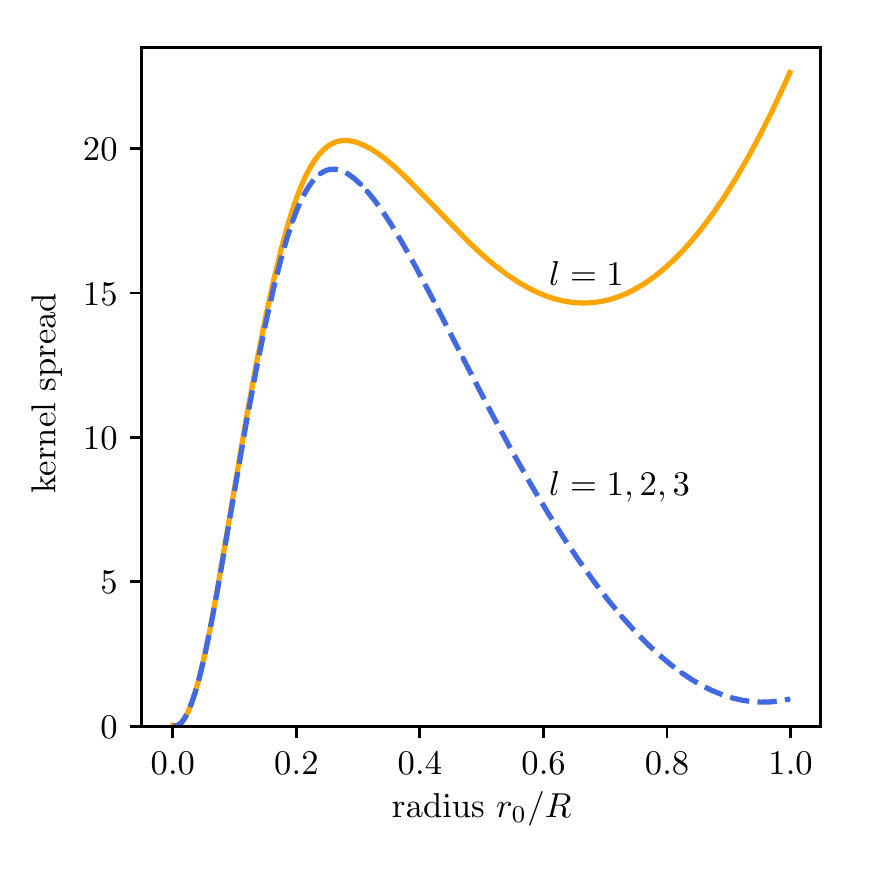}}
\caption{Dimensionless averaging-kernel spread (defined by the first part of Eq.~\ref{eqobj}) as a function of target radius. The averaging kernel spread for rotationally split dipole modes (orange solid line) and modes with $l\leq3$ (blue dashed line) is shown.}
\label{figspread}
\end{figure}

An alternative way to view the change in sensitivity between the mode sets is to investigate the spread of the averaging kernels. The averaging-kernel spread of the different mode sets is summarised in Fig.~\ref{figspread}. The kernel spread as given by the first term in Eq.~(\ref{eqobj}) is indicative of how well the respective averaging kernel is localised at the chosen target radius. The comparison shows that the averaging kernels in the core are well localised for both mode sets. However, the averaging kernels close to the surface are considerably more localised when $l=2$ and $l=3$ modes are added to the set of dipole modes.

For our analysis we assumed that the rotational splittings as given by Eq.~(\ref{eqsplitting}) have a linear dependence on the internal rotation profile $\Omega(r)$, resulting in splittings that are symmetric. However, \cite{deheuvels2017} demonstrated that the measured rotational splittings in the early red giant KIC~7341231 are not symmetric and depend non-linearly on the internal rotation profile. While this does not affect the results presented here as the synthetic splittings we use are symmetric by definition, the asymmetries in the rotational splittings have to be taken into account when rotational inversions are applied to observed rotational splittings of $l=2$ and 3 modes. In Appendix~\ref{secasym} we discuss the prospect of using observed asymmetric rotational splittings in linear rotational inversions.

\subsection{Rotation at intermediate radii}
\begin{figure}
\resizebox{\hsize}{!}{\includegraphics{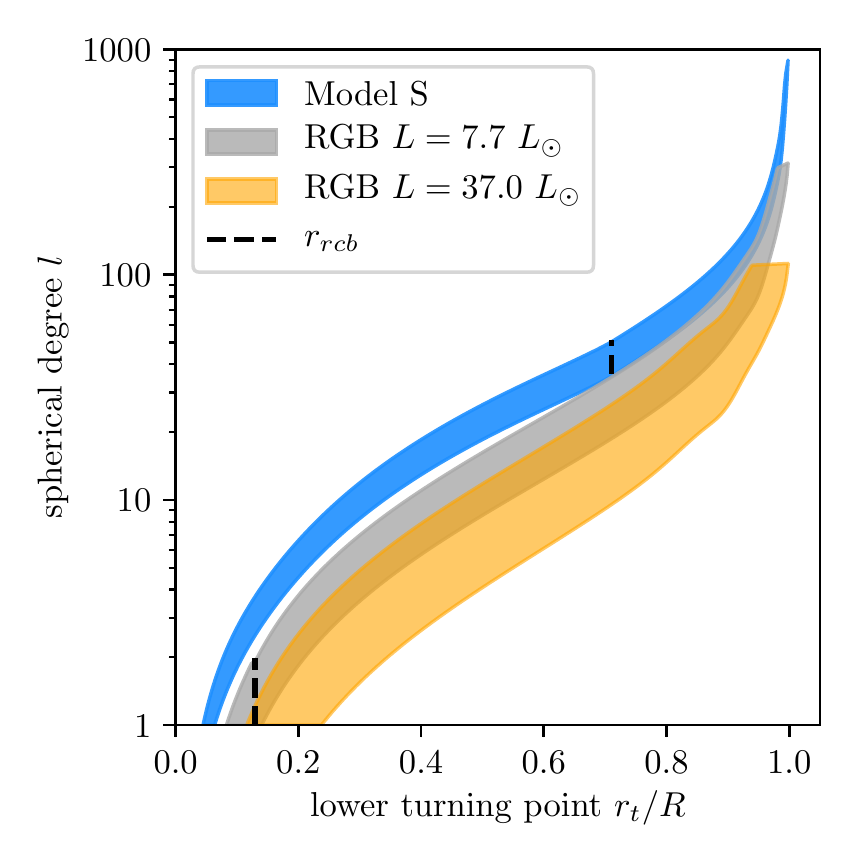}}
\caption{Spherical degree plotted as a function of the lower turning points of the oscillation modes. The shaded area indicates a range of $\pm4\Delta\nu$ around $\nu_\text{max}$ of the models indicated in the legend. The data for the red-giant model used in the inversions are shown in dark grey whereas the data for Model S are shown in blue. The more evolved red-giant model is shown in orange. The base of the convection zone is indicated with a dashed line. In the more evolved model the base of the convection zone is not visible on the given scale.}
\label{figlcurve}
\end{figure}
As shown in Fig.~\ref{figspread}, the localisation of averaging kernels at intermediate radii is not possible even when quadrupole and octopole modes are included. In order to do so, we extend our analysis to even higher spherical degrees. We look at the spherical degree of the oscillation modes as a function of the lower turning points. The lower turning point indicates where the oscillation mode spends a large fraction of its time, and hence is more sensitive to the internal rotation. In a mode set that includes modes with $l\leq l_\text{max}$, the localisation of averaging kernels will be possible for target radii lower than the lower turning point of the mode with spherical degree $l=l_\text{max}$.

We calculate the lower turning point as the intersection between the mode frequency and the Lamb frequency $S_l(r)$, where
\begin{align}
    S_l(r)=\frac{\sqrt{l(l+1)}\cdot c(r)}{r},
\end{align}
and $c(r)$ is the adiabatic sound speed \citep[e.g.][]{aerts2010}. This latter calculation can be applied to mixed modes with higher spherical degrees, as discussed below.

We calculated the lower turning points of modes with $l<1000$ for frequencies in the range {$\nu=\nu_\text{max}\pm4\Delta\nu$} as this is roughly the frequency range in which we expect to observe oscillation modes. The results for the low-luminosity red-giant model as described above are shown in Fig.~\ref{figlcurve}. As a comparison,  we also show the spherical degree as a function of lower turning point for the standard solar model known as Model S \citep{christensen1996} and a more evolved red-giant model in the same figure. Figure~\ref{figlcurve} shows that the spherical degree needed to recover the rotation rate at a given depth is lower for the red-giant model than for Model S and decreases further for a more evolved model along the RGB. To recover the rotation rate at an intermediate radius of $r/R=0.5$ a spherical degree of roughly $l\approx10$ is needed. Spherical degrees for selected values of lower turnings points for the low-luminosity red-giant model are listed in Table~\ref{tablelrt}. Figure~\ref{figlcurve} also suggests that it might be possible to probe the rotation around the base of the convection zone using only $l=1$ modes. However, our previous analysis has shown that this is in fact not the case. Here, we have to take into account that the dipole modes are mixed modes and thus have strong sensitivity to the deep core. This hinders the localisation of averaging kernels at the base of the convection zone. The coupling between the p- and the g-modes inside the star becomes weaker with increasing spherical degree. The strength of the coupling is also indicated by the radial distance between the lower turning point and the base of the convection zone in Fig.~\ref{figlcurve}. This means that although the p-dominated modes are mixed, they show less and less sensitivity to the deep layers with increasing spherical degree. As of $l=4$, the sensitivity in the core becomes negligible and the p-dominated modes act as pure p-modes. This shows that our approach of looking only at the Lamb frequency is justified, although this is only relevant to pure p-modes. Thus, it would be possible to measure rotation rates at intermediate radii in red giants using modes with a spherical degree of roughly $l\approx10$.

\begin{figure}
\resizebox{\hsize}{!}{\includegraphics{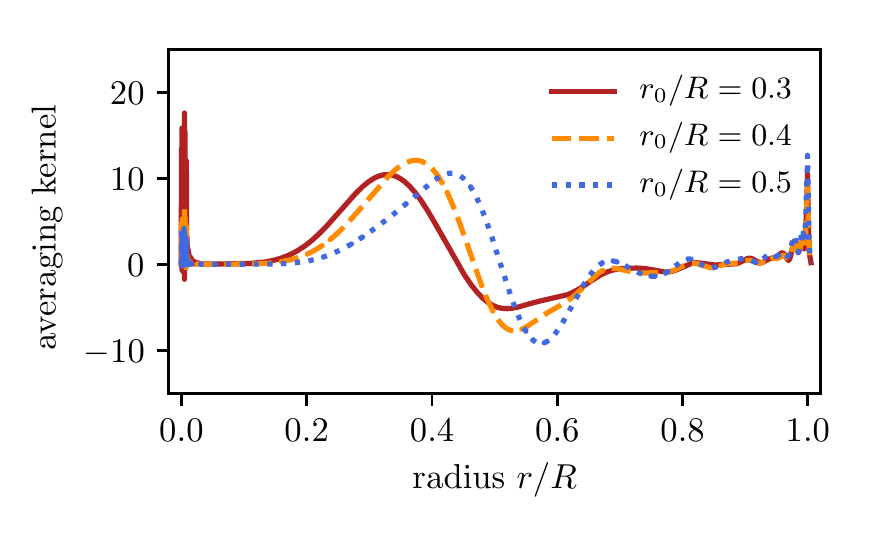}}
\caption{Averaging kernels for intermediate target radii calculated with a mode set containing 47 modes with $l\leq10$.}
\label{figinterm}
\end{figure}

\begin{figure}
\resizebox{\hsize}{!}{\includegraphics{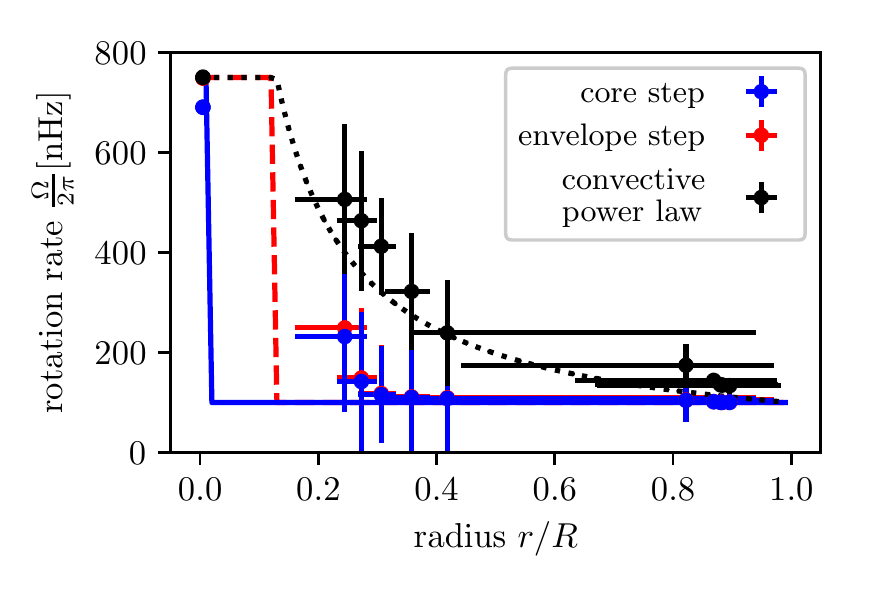}}
\caption{Synthetic and recovered rotation profiles. The profiles were calculated with rotational inversions on a mode set containing 47 modes with $l\leq10$.}
\label{figrotprofiles}
\end{figure}

\begin{table}
\caption{Spherical degrees for selected values of the lower turning point for the low-luminosity red-giant model as described above.}
\label{tablelrt}
\centering
\begin{tabular}{c c}
\hline\hline
\rule{0pt}{12pt} Lower turning point $r_t/R$ & Spherical degree $l$  \\[.5ex]
\hline
\rule{0pt}{10pt}0.3 & 6\\
0.5 & 13\\
0.7 & 26\\
0.9 & 72\\
\hline
\end{tabular}
\end{table}
In addition to our analysis of the lower turning points, we performed rotational inversions using a mode set with spherical degrees $l\leq10$. We extracted the rotational kernels from the same low-luminosity red-giant model as described above and selected 47 modes in the same frequency range as the $l\leq3$ mode sets. Rotational splittings with spherical degrees higher than $l=2$ have not yet been measured in red giants, and so we have no knowledge about the actual uncertainties on the rotational splittings; we therefore chose to apply the same uncertainty model as described in Appendix~\ref{secunc}. We again assumed that the rotational splittings are symmetric by calculating them using Eq.~(\ref{eqsplitting}). The averaging kernels at intermediate radii computed from this extended mode set are shown in Fig.~\ref{figinterm}. In contrast to the previously studied mode sets with spherical degree $l\leq3$, the averaging kernels in Fig.~\ref{figinterm} show sensitivity at intermediate radii. The peak sensitivity of each kernel is located close to the chosen target radius which means that the averaging kernels are localised at the specified radius. However, averaging kernels with target radii $r_0/R>0.5$  are not localised. This agrees well with the conclusions drawn from the analysis of the lower turning points of the oscillation modes.
We also used the extended mode set to recover the synthetic rotation profiles described in Appendix~\ref{secrotprofiles}. The resulting rotation profiles and the synthetic input profiles are shown in Fig.~\ref{figrotprofiles}. We plotted the estimated rotation rates against the median radius of the averaging kernels, which we define as the radius such that 50\% of the sensitivity lies below it and 50\% above it. The estimated rotation rates recover the shape of the synthetic rotation profiles for radii $r/R\geq0.2$. Between 0.005 and 0.2, the localisation of averaging kernels was not possible, indicating that there is not enough sensitivity to this region in the mode set. The core rotation rate is recovered quite accurately in case of the envelope step and convective power law profile, which is similar to our results for the mode sets with $l\leq3$ (see Table~\ref{tablel1}, lower part). Still, the estimated core rotation rate for the core step profile deviates substantially from the input. This supports our previous conclusion that modes with $l\leq3$ are the only source of information for the core rotation in red giants. The offset between the recovered and the input rotation rates at intermediate radii can be explained by the sensitivity of the corresponding averaging kernels to the core region. The error bars on the estimated rotation rates are small enough that it is possible to distinguish between the convective power law profile and a step-like profile from the asteroseismic estimate alone. Thus, our analysis shows that mode sets with spherical degrees of $l_\text{max}\approx 10$ enable us to measure rotation rates at intermediate radii.
\section{Conclusion}
We performed rotational inversions for a series of models with masses of 1.0, 1.5, and 2.0~$M_\odot$ that span evolutionary phases from the base of the RGB up to the bump in luminosity. We studied the theoretical possibility of estimating internal rotation rates in RGB stars using different mode sets with spherical degrees from $l=1$ up to $l=10$. Our study shows how the results of rotational inversions using dipole ($l=1$) modes vary as the stellar models evolve along the RGB and that they are essentially insensitive to stellar mass and metallicity. For all models under consideration it was possible to compute well localised core-averaging kernels and to probe mean core rotation rates confirming results from other studies \cite[e.g.][]{deheuvels2014,mosser2012}. The grey symbols in Fig.~\ref{figsensMESA} show that more than 99\% of the sensitivity is confined below the base of the convection zone in all evolutionary stages. As a consequence mean core rotation rates can be determined with nearly no systematic errors (see Appendix~\ref{sectradeoff} for the definition of systematic errors). However, surface-averaging kernels behave quite differently. At the base of the RGB, the surface-averaging kernel is localised to about 95\% in the convective envelope.  Consequently, the estimated surface rotation rates deviate from the underlying rotation rate for the $l=1$ mode set. This is due to the large spread of the surface-averaging kernel and its sensitivity to the stellar core. Hence the estimated surface rotation rates using the $l=1$ mode set are only upper limits on the true underlying surface rotation. This is in line with results from previous studies \citep{deheuvels2012,deheuvels2014,dimauro2016}. As the stellar models evolve up the giant branch, the surface-averaging kernel becomes more and more sensitive to the stellar core until the surface sensitivity reaches a minimum of about 70\%. Interestingly, the sensitivity to the surface rotation starts to increase again as the models evolve even further. The models reach another maximum in surface sensitivity just below the red-giant bump. The surface sensitivity is comparable to the models at the base of the RGB. This behaviour is common to all masses and metallicities considered. This indicates that the systematic errors on estimates of the average rotation rate of the envelope should be the same at the base of the RGB and just below the red-giant bump. As a possible explanation for the reduced surface sensitivity, we suggest the glitch in the buoyancy frequency. As stars evolve along the RGB the convective envelope starts to recede at some point. This leaves behind a chemical discontinuity which imprints a glitch in the buoyancy frequency. This glitch moves inward in radius as the models evolve further up the RGB. As long as the glitch lies within the frequency range of the modes used, the sensitivity to the surface rotation is decreased. When the glitch moves out of the frequency range, the coupling decreases and modes become more sensitive to the surface again. 
We  artificially removed the glitch in several models and found that the minimum in surface sensitivities disappeared.

Our analysis of one low-luminosity red-giant model shows that low-degree ($l\leq3$) modes probe mean core rotation rates. Using dipole modes it is possible to form averaging kernels that are nearly completely confined below the convection zone ($r<r_\text{rcb}$) \citep[see ][]{deheuvels2012,deheuvels2014,dimauro2016}.  These averaging kernels can be used to estimate the mean rotation rate of the star below the convection zone. However, the estimated core rotation rate is still significantly different from the true rate in the case of the core step rotation profile due to the sensitivity of the core-averaging kernel to the lower rotation rate outside $1.5r_\text{H}$. The localisation of the core-averaging kernel only marginally improved when modes of $l=2,3$ were added, as they are mostly sensitive to the envelope of the star. For higher spherical degrees, the coupling between the p- and the g-modes becomes even weaker, which makes g-dominated mixed modes harder to observe. The low-degree modes therefore remain the only current source of information about the deep core of red-giant stars.

In comparison to the localisation of the core-averaging kernels, the localisation of the surface-averaging kernels is considerably improved when using modes of $l=2,3$ compared to only using $l=1$ modes. This is in line with results presented in \cite{dimauro2018}. The sensitivity of the surface-averaging kernel is almost completely confined above the base of the convection zone ($r>r_\text{rcb}$). Here, the $l=3$ modes would be sufficient to get the same localisation of the surface-averaging kernel as they show most of their sensitivity towards the surface. However, the estimated rotation rate is still not exclusively sensitive to the stellar surface. Rather, it provides a mean rotation rate of the stellar envelope independent of the core rotation. This makes it possible to distinguish between solid body and differential rotation in the convection zone when another independent measure of the surface rotation --for example spot modulation or spectroscopic measurements-- is available. When the asteroseismically determined surface rotation rate is larger than the independent measurement from spectroscopy or spot modulation, we can conclude that the star rotates differentially in the convection zone with a rotation rate that increases with depth. When the asteroseismic surface rotation rate is roughly equal to the independent measurement of the surface rotation, the star rotates as a solid body in the convection zone. \cite{beck2018} applied the above approach to the primary component of the binary KIC~9163796 and showed that the amount of differential rotation in the convection zone is small, meaning that the convection zone is nearly rotating as a solid body. This could provide important hints for underlying angular momentum transport mechanisms. In the case of the solid body rotation in the convection zone, core and envelope rotation seem to be decoupled. On the other hand, differential rotation in the convection zone points towards angular momentum transport from the core to the envelope and a coupling between core and envelope rotation \citep{kissin2015}.

The study of higher degree modes currently remains theoretical because rotationally split $l>2$ modes have not  yet been detected. Modes with $l=3$ are in principle observable using measurements of the integrated stellar disc. To disentangle the rotational splitting from the line width, a high resolution of frequency is required, together with low noise. To achieve this, long and uninterrupted observations of stellar brightness fluctuations with a length on the order of years---as for example obtained with the  \textit{Kepler} telescope---are necessary. Although rotationally split   $l=3$ modes have not yet been detected in the \textit{Kepler} data, further analysis of the data could still reveal rotationally split $l=3$ modes. As shown above, a small number of $l=3$ modes is  sufficient to improve the inversion results compared to dipole modes only. The $l\leq 3$ mode set discussed in Sect.~\ref{secres} contains four $l=3$ modes. The detection of $l=3$ modes is easier in radial velocity measurements than in photometry due to a decreased signal of the granulation background \citep{grundahl2007}. For this reason, the Stellar Observations Network Group (SONG), which is building a network of 1~m telescopes equipped with high-resolution spectrographs, is expected to improve our capacity to detect these modes. \cite{grundahl2017} measured the rotational splitting of dipole modes in the subgiant star $\mu$~Herculis using the first node of the SONG network. In addition, these latter authors detected a number of $l=3$ modes, although no measurement of a rotational splitting was reported.

The analysis of modes with $l>3$ showed that modes with a spherical degree of $l\approx10$ might be sufficient to recover rotation rates at intermediate radii. This would be beneficial for setting more constraints on angular momentum transport and rotational mixing operating at that depth in the stellar interior. With current methods, modes with spherical degrees larger than three are not observable due to cancellation effects in the integrated light. New methods relying on interferometric techniques \citep[for a review see][]{cunha2007} may be able to recover modes with higher spherical degrees to probe rotation at intermediate radii.
\label{seccon}
\begin{acknowledgements}
The research leading to the presented results has received funding from the European Research Council under the European Community’s Seventh Framework Programme (FP7/2007-2013)/ERC grant agreement no 338251 (StellarAges). SB acknowledges partial funding from NASA grant NNX16A109G and NSF grant AST-1514676.
\end{acknowledgements}

\bibliographystyle{aa}
\bibliography{bibliography}
\begin{appendix}
\section{Synthetic data}
\label{secdataapp}
\subsection{Stellar models}
For the MESA models we used the solar mixture of heavy elements as derived by \cite{grevesse1993}. We applied OPAL opacities \citep{iglesias1996} and the OPAL equation of state \citep{rogers2002}. We used an Eddington grey atmosphere \citep[][]{eddington1926}. Convection was treated according to the mixing length theory \citep{boehm1958}. We set the mixing length parameter to a solar calibrated value of $\alpha_{\text{MLT},\odot}=1.67$. We also computed models with different mixing length parameters and different atmospheres and found that our conclusions are not influenced by the choice of these parameters. The mass fractions $X,Z,Y$ for the different metallicities were computed in a consistent way assuming a linear helium--metallicity relation with $Y_\text{p}=0.2463$ \citep{coc2014} and $\frac{\Delta Y}{\Delta Z}=0.9945$ from a solar model without elemental diffusion. 
The same opacities, mixture of heavy elements, and atmosphere were used in the GARSTEC models. Convection was treated according to the mixing length theory using a solar calibrated mixing length parameter $\alpha_{\text{MLT},\odot}=1.58$.
\subsection{Synthetic rotation profile}
\label{secrotprofiles}
To study the results of the rotational inversions, we use synthetic rotation profiles based on the following scheme from \cite{klion2017}:
\begin{align}
    \Omega(r)=\left\{
                \begin{array}{ll}
                  \Omega_\text{c}\,\,\,\,\,\,\,\,\,\text{for}\,\,\,\,\,\,\, r\leq 1.5r_\text{H}\\[5pt]
                  \Omega_\text{m}\,\,\,\,\,\,\,\,\,\text{for} \,\,\,\,\,\,\, 1.5r_\text{H}< r\leq r_\text{rcb}\\[5pt]
                  \Omega_\text{e}\cdot\left (\frac{R}{r}\right )^\gamma\,\,\,\,\,\,\,\,\,\text{otherwise,}
                \end{array}
              \right.
\end{align}
where $r_\text{H}$ indicates the radius of the hydrogen burning shell, and $r_\text{rcb}$ is the boundary between the radiative and convective zones. The power-law decrease in the convection zone was derived based on theoretical grounds by \cite{kissin2015}, who showed that $\gamma$ takes values between 1 and 1.5. To get a continuous rotation profile, \cite{klion2017} set 
\begin{align}
\gamma=\dfrac{\log\left(\dfrac{\Omega_\text{m}}{\Omega_\text{e}}\right)}{\log\left(\dfrac{R}{r_\text{rcb}}\right)}\,\,.
\label{eqgamma}
\end{align}

\begin{figure}
\resizebox{\hsize}{!}{\includegraphics{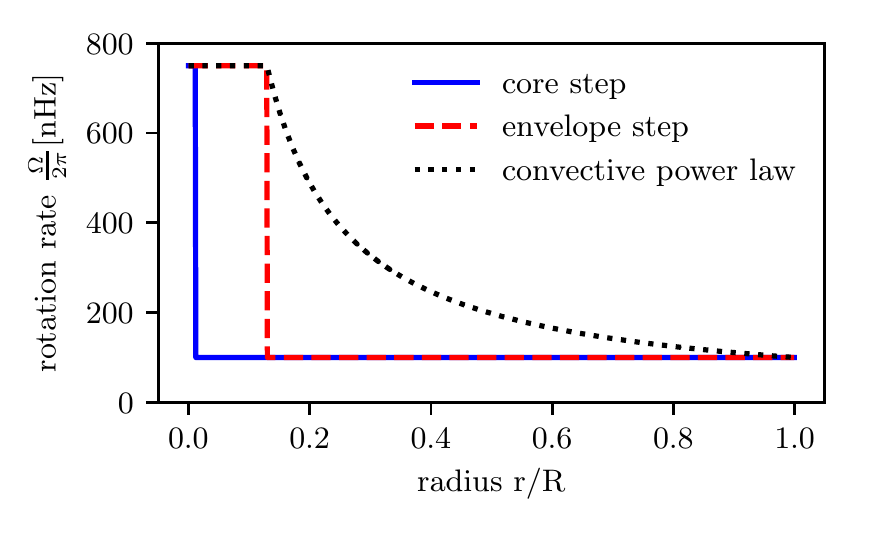}}
\caption{Synthetic rotation profiles with a step just outside the hydrogen-burning shell (core step, blue solid line), a step at the base of the convection zone (envelope step, red dashed line) and power law decrease in the convection zone (convective power law, black dotted line). }
\label{figprofiles}
\end{figure}

For our analysis, we used three different synthetic rotation profiles. In the first rotation profile the star rotates with a constant rotation rate $\Omega_\text{c}$ below $1.5r_\text{H}$ and outside 1.5$r_\text{H}$ with another rotation rate ($\Omega_\text{m}=\Omega_\text{e}$) which implies $\gamma=0$. We refer to this profile as the `core step profile'. In the second synthetic profile the star rotates with a constant rotation rate ($\Omega_\text{m}=\Omega_\text{c}$) below the convection zone and outside with another constant rotation rate $\Omega_\text{e}$ (where $\gamma=0$). In this way, the rotation profile has a step at the boundary between the radiative and convective zones $r_\text{rcb}$. We refer to this profile as the `envelope step profile'. We have chosen the third profile such that the star rotates with a single rotation rate below the convection zone ($\Omega_\text{m}=\Omega_\text{c}$) and used the formulation for $\gamma$ from \cite{klion2017} (Eq.~\ref{eqgamma}) in the convection zone. In the following, we refer to this profile as the `convective power law profile'. Following \cite{dimauro2016} we have chosen $\Omega_\text{c}=750$~nHz and $\Omega_\text{e}=100$~nHz, which is close to observed values in the low-luminosity red giant KIC~4448777. We determined the values for $r_\text{H}$ and $r_\text{rcb}$ using the stellar model profiles. The three different rotation profiles are shown in Fig.~\ref{figprofiles}.

The synthetic rotation profiles are used to calculate synthetic rotational splittings as the input for the rotational inversions using Eq.~(\ref{eqsplitting}) and rotational kernels from the red-giant models. We note that the formation of averaging kernels is independent of the input rotation profile.

\subsection{Mode sets}
\label{secmodesets}
\begin{figure}
\resizebox{\hsize}{!}{\includegraphics{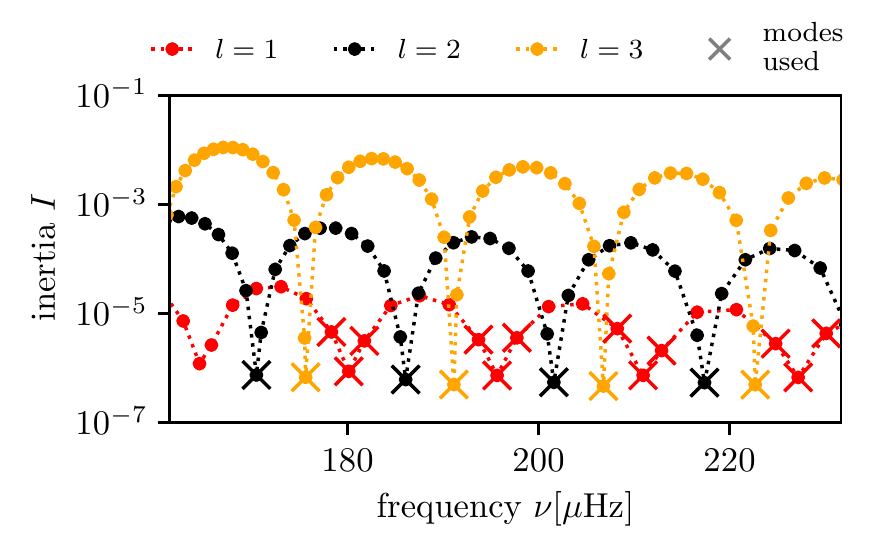}}
\caption{Mode inertia of modes with spherical degrees 1 (red), 2 (black), and 3 (orange). The modes which were selected are marked with crosses.}
\label{figinertia}
\end{figure}

For the rotational inversions along the lower RGB, we used dipole modes only. The mode set contains 12 dipole modes in a frequency range of approximately $\nu_\text{max}\pm2\Delta\nu$. This set contains the three dipole mixed modes per acoustic radial order with lowest inertia. These are the most p-dominated and two of the least g-dominated modes. The modes used are indicated with red crosses in Fig.~\ref{figinertia} for an example stellar model. This mode set is comparable to observed mode sets in red giants and we refer to it as the $l=1$ mode set. 

For the detailed study of rotational inversion results in a single red-giant model, we defined a second mode set. The mode set includes pressure dominated quadrupole ($l=2$) and octopole ($l=3$) modes in a frequency range of $\nu_\text{max}\pm2\Delta\nu$. As the visibility of the modes decreases with increasing inertia we stick to using only the $l=2$ and 3 modes that have the lowest inertia. According to the frequency range $\nu_\text{max}\pm2\Delta\nu,$ this mode set contains 4 quadrupole and 4 octopole modes 
in addition to the 12 dipole modes, making 20 modes in total (see~Fig.~\ref{figinertia}). We refer to this mode set as the $l=1,2,3$ mode set.
\subsection{Model of uncertainties of the splittings}
\label{secunc}
To interpret the results of the rotational inversions for synthetic data, it is important to have a meaningful model for the uncertainties of the synthetic rotational splittings. Following \cite{schunker2016a}, we fit a quadratic function to the uncertainties of the splittings of KIC 4448777 derived with an MCMC algorithm. This function is used to associate uncertainties with the synthetic rotational splittings. We would like to point out that while the synthetic splittings are exact, realistic uncertainties are needed so that the inversion results of the synthetic data can be used to determine properties of inversion results obtained with real data. To compute synthetic splitting uncertainties for the stellar models, the quadratic function was centred around the model value of $\nu_\text{max}$. It was then skewed or stretched to achieve a fixed reference uncertainty of $\sigma_\text{ref}=0.007\mu\text{Hz}$ at a frequency of $\nu_\text{max}\pm\Delta\nu$. To facilitate the interpretation of the inversion results we use noise-free synthetic rotational splittings as an input for the rotational inversions as suggested by \cite{christensen1990}. We note that the localisation of the averaging kernels does not depend on the values of the rotational splittings, but only on the uncertainty model used.\\
\section{Choice of trade-off parameter}
\label{sectradeoff}
\begin{figure}
\resizebox{\hsize}{!}{\includegraphics{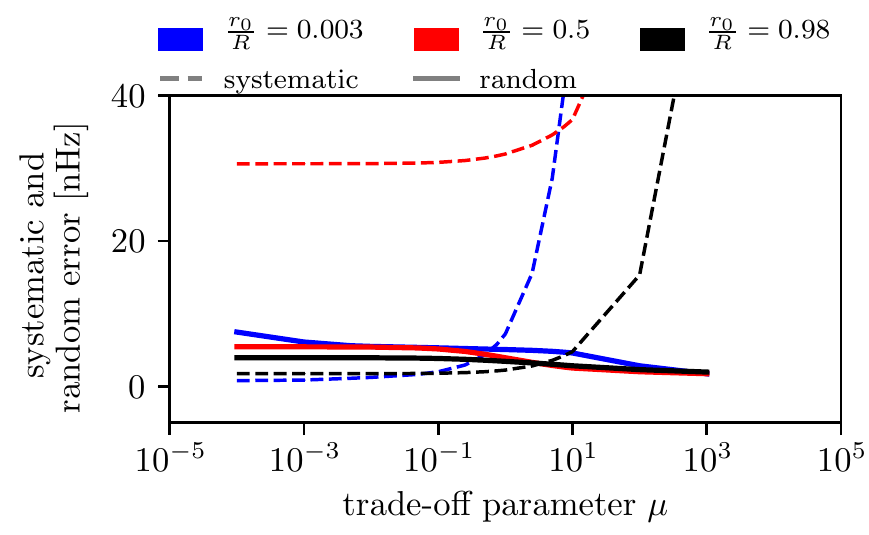}}
\caption{Systematic and random errors as a function of the trade-off parameter for the rotational inversion using the $l=1,2,3$ mode set as described in Sect. \ref{secmodesets}.}
\label{figtradeoff}
\end{figure}

As described in Sect.~\ref{secrotinv} the inversion procedure consists of a trade-off parameter which regulates between the localisation of the averaging kernels and the propagation of data uncertainties, i.e. random errors. In order to obtain localised results with meaningful uncertainties, the trade-off parameter must be chosen properly. We chose the trade-off parameter by comparing random and systematic errors of the inversion results for different values of the trade-off parameter. The random errors in the inferred rotation rates are derived from an error propagation of the uncertainties on the rotational splittings. The systematic errors are defined as the absolute difference between the input rotation rate and the recovered rotation rate at a given radius. In this study, it is possible to access the systematic errors because the underlying rotation profile is known. We assumed that the trade-off parameter is optimal when random and systematic errors are equal. For higher trade-off parameters, the systematic errors would exceed the random errors, making the random errors meaningless. For lower trade-off parameters, the random errors are potentially overestimated with respect to the optimal trade-off parameter. The random and systematic errors for the $l=1,2,3$ mode set, as described in Sect.~\ref{secmodesets}, are shown in Fig.~\ref{figtradeoff}. The curves of random and systematic errors for target radii of $r_0/R=0.003$ and $r_0/R=0.98$ do intersect, which shows that the systematic errors are rather low. For a target radius of $r_0/R=0.5$, the systematic errors are higher than the random errors even for $\mu=0$. To ensure that the systematic errors are lower than the random errors for all target radii where this is possible, we chose the smallest of the intersection trade-off parameters, which in this case is the value for $r_0/R=0.003$. The optimal trade-off parameters determined for the two mode sets as defined in Sect.~\ref{secmodesets} are listed in Table~\ref{tabletradeoff} for the low-luminosity red-giant model as described in Sect.~\ref{secres}.

\begin{table}
\caption{Optimal trade-off parameters for the mode sets described in Sect.~\ref{secmodesets}.}
\label{tabletradeoff}
\centering
\begin{tabular}{c c}
\hline\hline
\rule{0pt}{12pt} Mode set & Trade-off parameter $\mu$  \\[.5ex]
\hline
\rule{0pt}{10pt}$l=1$  & 0.42\\
$l=1,2,3$ & 0.53\\
\hline
\end{tabular}
\end{table}
\section{Asymmetric rotational splittings}
\label{secasym}
\cite{deheuvels2017} studied the effects of near degeneracy effects on rotationally split $l=2$ modes in the early red giant KIC 7341231. These latter authors measured the frequencies of two rotationally split and near degenerate $l=2$ modes and showed that they are significantly asymmetric.  When the frequency difference between two modes with the same spherical degree becomes comparable to the rotation rate of the star $\left(|\omega_\text{a}-\omega_\text{b}|\sim\Omega\right)$ near degeneracy effects come into play. In the presence of near degeneracy effects the rotational splittings are no longer symmetric and Eq.~(\ref{eqsplitting}) becomes invalid. Instead the perturbed frequencies are given by:
\begin{align}
\omega_\pm=\frac{\omega_\text{a}+\omega_\text{b}}{2}\pm\frac{1}{2}\sqrt{(\omega_\text{a}-\omega_\text{b})^2-4\delta\omega_\text{ab}^2},
\end{align}
with the symmetrically perturbed frequencies $\omega_\text{a}=\omega_{0,\text{a}}+m\,\delta\omega_\text{a}$ and for $\omega_\text{b}$ respectively. The term $\delta\omega_\text{ab}$ describes the interaction between both modes and is given by:
\begin{align}
\delta\omega_\text{ab}=m\int_0^R\mathcal{K}_\text{ab}(r)\Omega(r)\,\text{d} r.
\end{align}
The function $\mathcal{K}_{ab}$ is composed of the radial and horizontal eigenfunctions $\xi_\text{r}$ and $\xi_\text{h}$ of the two involved modes: 
\begin{equation}
\begin{split}
        \mathcal{K}_\text{ab}=\rho_0r^2&\left[\xi_{\text{r},0,\text{a}}\xi_{\text{r},0,\text{b}}+(L^2-1)\xi_{\text{h},0,\text{a}}\xi_{\text{h},0,\text{b}}\right.\\
        &\left.-\xi_{\text{r},0,\text{a}}\xi_{\text{h},0,\text{b}}-\xi_{\text{h},0,\text{a}}\xi_{\text{r},0,\text{b}}\right]
    \end{split}
\end{equation}
with $L^2=l(l+1)$.

\begin{figure}
\resizebox{\hsize}{!}{\includegraphics{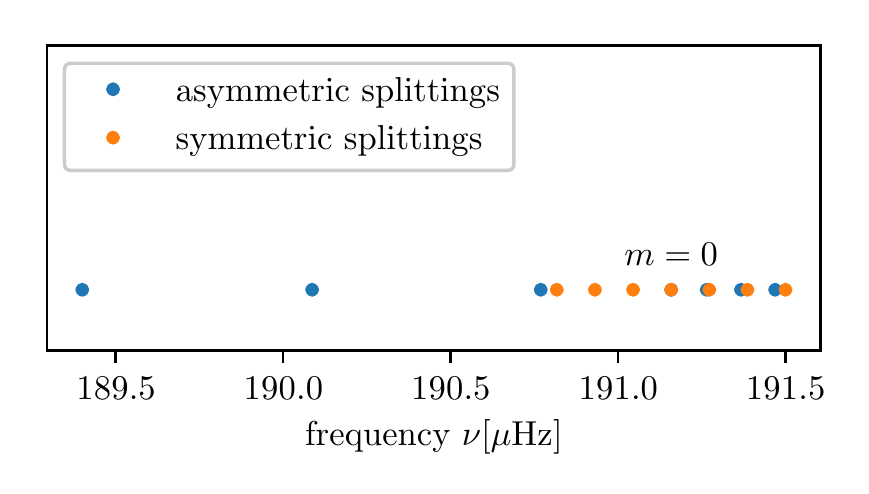}}
\caption{Rotationally split $l=3$ p-dominated mode with a central frequency of $\omega_\text{b}/(2\pi)=191.16\mu$Hz. Multiplet components calculated without near degeneracy effects are shown in orange, splittings calculated including near degeneracy effects are shown in blue.}
\label{figasymm}
\end{figure}

Using this formulation we calculated asymmetric rotational splittings for a p-dominated $l=3$ mode. The p-dominated mode has a frequency of $\nu_\text{b}=191.16\mu$Hz. The close by g-dominated mode has a higher frequency of $\nu_\text{a}=191.47\mu$Hz. The frequency difference of $0.31\mu$Hz is comparable to observed rotation rates in red giants which are on the order of 0.1 to $1\mu$Hz. The resulting frequencies for all seven components are shown in Fig.~\ref{figasymm}. Figure~\ref{figasymm} shows that the components with $m<0$ are strongly effected by the higher frequency g-dominated mode and the rotational splitting increased substantially compared to the symmetric case. However, for the components with $m>0$ the frequencies deviate much less from the symmetric case. The deviations are on the same order as typical uncertainties of observed $l=1$ rotational splittings. This means that they follow the non-degenerate description within the uncertainties and can be used in linear rotational inversions in the way described above.
\section{Additional results}
\begin{figure}
\resizebox{\hsize}{!}{\includegraphics{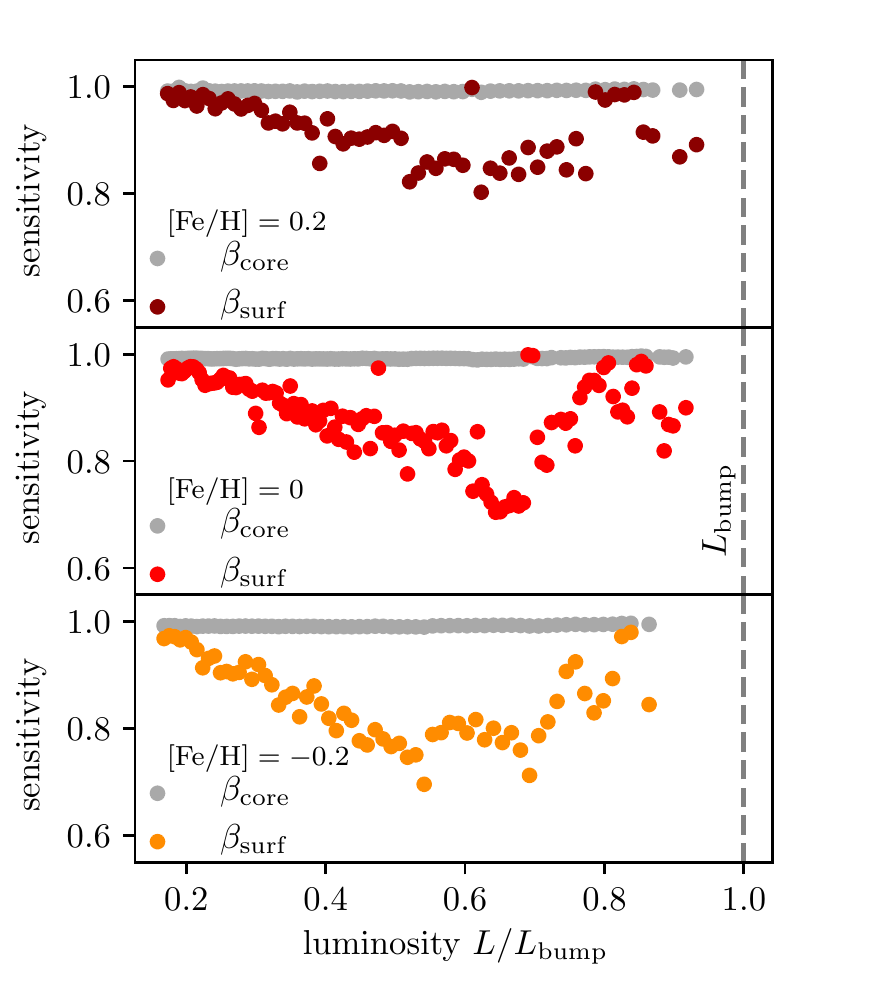}}
\caption{Core and surface sensitivities $\beta_\text{core}$ and $\beta_\text{surf}$ for the $1~M_\odot$ evolutionary tracks with $[\text{Fe}/\text{H}]=-0.2,0$ and 0.2.}
\label{figsensmetal}
\end{figure}

\begin{figure}
\resizebox{\hsize}{!}{\includegraphics{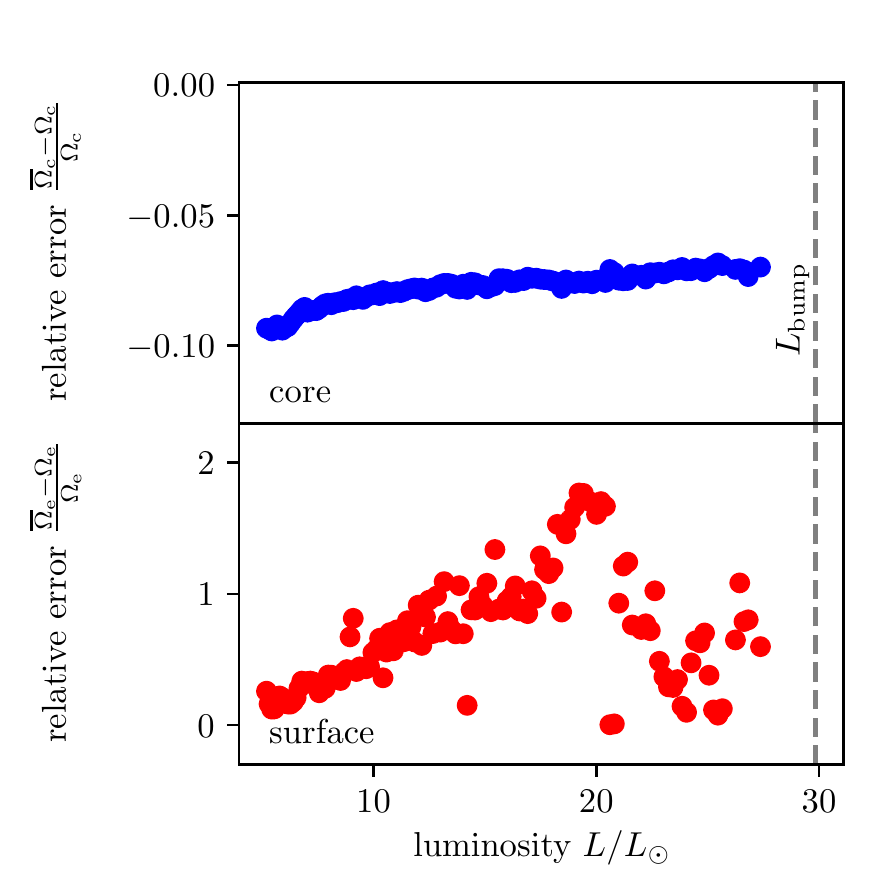}}
\caption{Relative errors of core and surface rotation rates along the RGB for $1~M_\odot$, $[\text{Fe}/\text{H}]=0$ MESA models. Here the core step profile has been used.}
\label{figrelerrcorestep}
\end{figure}

\begin{figure}
\resizebox{\hsize}{!}{\includegraphics{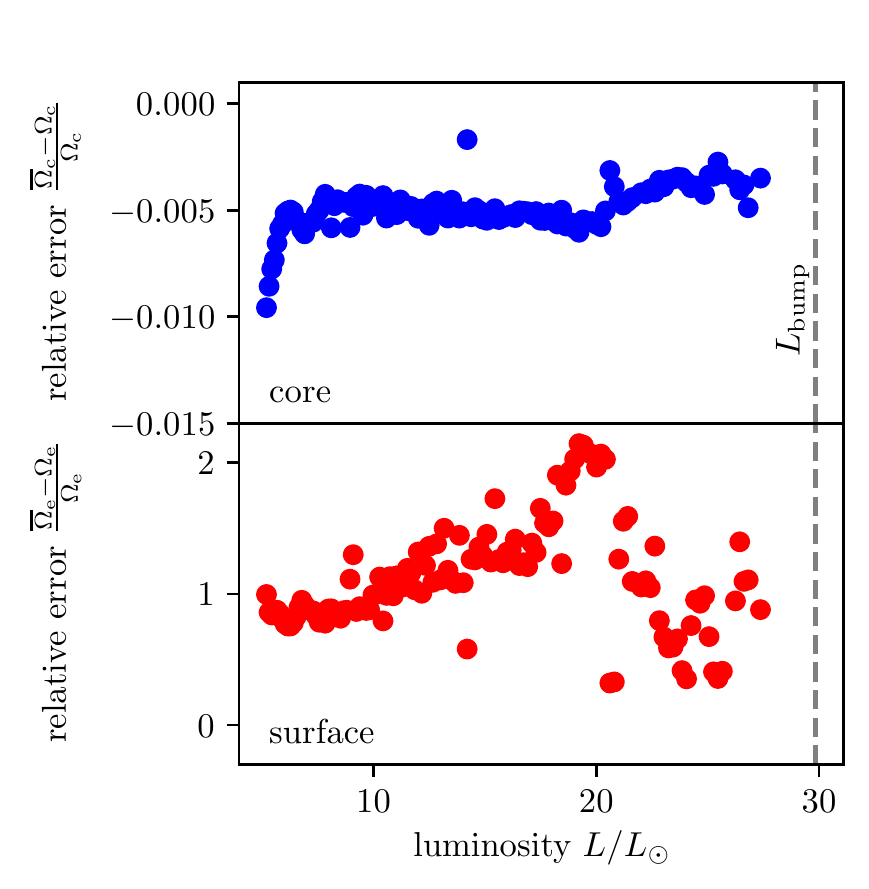}}
\caption{Relative errors of core and surface rotation rates along the RGB for the $1~M_\odot$, $[\text{Fe}/\text{H}]=0$ MESA models. Here the convective power law profile has been used.}
\label{figrelerrrpowerlaw}
\end{figure}

\begin{figure}
\resizebox{\hsize}{!}{\includegraphics{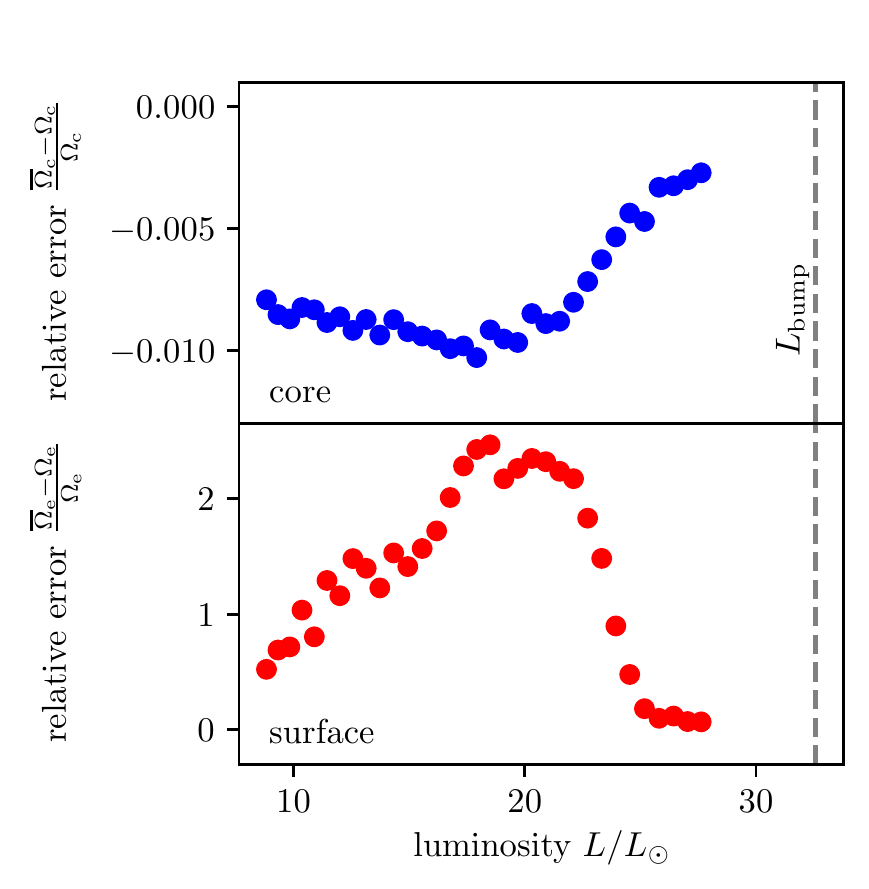}}
\caption{Relative errors of core and surface rotation rates along the RGB for the $1~M_\odot$, $[\text{Fe}/\text{H}]=0$ GARSTEC models. Here the envelope step profile has been used.}
\label{figinvrgbGARS}
\end{figure}

Figure~\ref{figsensmetal} shows the core- and surface-sensitivities $\beta_\text{core}$ and $\beta_\text{surf}$ for the 1~$M_\odot$ MESA models with, $[\text{Fe}/\text{H}]=0.2,0$ and $-0.2$ in the upper, middle, and lower panel respectively. As already discussed in Sect.~\ref{secinvRGB} for the $[\text{Fe}/\text{H}]=0$ models also the higher and lower metallicity models show a high sensitivity to the core rotation along the RGB. The surface sensitivities exhibit a minimum at about $0.6-0.8~L_\text{bump}$. Thereafter the surface sensitivity increases again.\\
Figure~\ref{figrelerrcorestep} shows the relative errors of the core and surface rotation rates for the $1.0~M_\odot, [\text{Fe}/\text{H}]=0$ MESA models when the core step profile described in Appendix~\ref{secrotprofiles} is used. The relative errors for the core rotation rates differ substantially from the relative errors obtained with the envelope step profile presented in Fig.~\ref{figinvrgbMESA} (upper panel). This can be explained by the leakage of the core-averaging kernels outside the fast rotating region as described in Sect.~\ref{secres} for the low-luminosity red-giant model. The relative errors for the estimated surface rotation rates using the core step profile appear very similar to the relative errors obtained with the envelope step profile presented in Fig.~\ref{figinvrgbMESA} (lower panel). This is the case because the surface averaging kernels do not have substantial sensitivity to the rotation in the range $1.5r_\text{H}$ to $r_\text{rcb}$ in which the rotation rate differs.\\
Figure~\ref{figrelerrrpowerlaw} shows the relative errors of the core and surface rotation rates for the $1.0~M_\odot, [\text{Fe}/\text{H}]=0$ MESA models when the convective power law profile described in Appendix~\ref{secrotprofiles} is used. In this case the relative errors for the core rotation rates look similar to the results in Fig.~\ref{figinvrgbMESA} (upper panel) as the core-averaging kernels do not leak outside the fast rotating region substantially. However, the relative errors for the surface rotation rates differ from the results shown in Fig.~\ref{figinvrgbMESA} (lower panel). Due to the large spread of the surface-averaging kernels, the estimated surface rotation rates pick up some of the rotation in the convective envelope, which rotates faster in the convective power law profile than in the envelope step profile. This increases the relative errors of the estimated surface rotation rates for all models. For a more detailed description of the effect of a different underlying rotation profile we refer again to Sect.~\ref{secres}.\\
Figure~\ref{figinvrgbGARS} shows the relative errors for core and surface rotation rates using the envelope step rotation profile for the GARSTEC $1.0~M_\odot, [\text{Fe}/\text{H}]=0$ models. They closely resemble the results obtained with the MESA models shown in Fig.~\ref{figinvrgbMESA}.
\end{appendix}
\end{document}